%% file: ddt-paper-a.tex
\documentclass[useAMS,usenatbib]{mn2e}

\RequirePackage{ifpdf}

\usepackage[T1]{fontenc}
\usepackage[utf8]{inputenc}

\usepackage{amsmath,amsfonts,amssymb}
\usepackage{xspace, lastpage}
\usepackage{color,graphicx}
\graphicspath{{./}{./fig/}}

\usepackage[mediumspace,mediumqspace,Grey,squaren]{SIunits}

\definecolor{LinkColor}{rgb}{0.0,0.2,0.5}
\usepackage[pdfpagemode=UseNone,pdfstartview=FitB]{hyperref}
\hypersetup{
  breaklinks=true,
  colorlinks=true,
  linkcolor=LinkColor,
  citecolor=LinkColor,% ForestGreen,% default: green
  filecolor=LinkColor,% Purple,% default: magenta
  menucolor=LinkColor,% BrickRed,% default: red
  pagecolor=LinkColor,% BrickRed,% default: red
  urlcolor=LinkColor,% default: cyan
}

\input{ddt-macros}

\definecolor{DarkOrange}{rgb}{0.6,0.6,0.0}
\definecolor{DarkRed}{rgb}{0.7,0.0,0.0}
\definecolor{DarkGreen}{rgb}{0.3,0.7,0.0}
\definecolor{DarkBlue}{rgb}{0.2,0.3,0.7}
\definecolor{FireBrick}{rgb}{0.70,0.13,0.13}
\definecolor{Yellow}{rgb}{0.95,0.95,0.00}
\definecolor{Black}{rgb}{0.0,0.0,0.0}

\title[3-D deconvolution of hyper-spectral astronomical data]%
      {3-D deconvolution of hyper-spectral astronomical data}

\author[S.\ Bongard,  F.\ Soulez, \'E.\ Thi\'ebaut,  \'E.\ Pecontal]%
       {S.\ Bongard$^{1}$\thanks{E-mail:bongard@in2p3.fr},
	 F.\ Soulez$^{2,3}$,
	 \'E.\ Thi\'ebaut$^{2}$, and
	 \'E.\ Pecontal$^{2}$\\
${}^{1}$Laboratoire de Physique Nucl\'eaire et des Hautes \'Energies,
         Universit\'e Pierre et Marie Curie Paris 6, \\
$\phantom{{}^{1}}$Universit\'e Paris
      Diderot Paris 7, CNRS-IN2P3, 4 place Jussieu, 75252 Paris Cedex 05, France\\
${}^{2}$Université de Lyon, Lyon, F-69003, France; Université Lyon
  1, Observatoire de Lyon, 9 avenue Charles André, \\
$\phantom{{}^{2}}$Saint-Genis Laval, F-69230, France; CNRS, UMR 5574,
  Centre de Recherche Astrophysique de Lyon; \'Ecole \\
$\phantom{{}^{2}}$Normale Supérieure de Lyon, Lyon, F-69007, France.\\
${}^{3}$Centre Commun de Quantim\'etrie, Université Lyon 1, 8 avenue Rockefeller, 69373 Lyon cedex 08, France.}

\begin{document}

\date{Accepted 2011 July 20.  Received 2011 July 20; in original form 2011 March 25}

\pagerange{\pageref{firstpage}--\pageref{lastpage}} \pubyear{}

\maketitle

\label{firstpage}

%------------------------------------------------------------------- ABSTRACT -
\begin{abstract}
  In this paper we present a general method for multichannel image restoration
  based on regularized $\chi^{2}$. We introduce separable regularizations that
  account for the dynamic of the model and take advantage of the continuities
  present in the data, leaving only two hyper-parameters to tune.

  We illustrate a practical implementation of this method in the context of
  host galaxy subtraction for the Nearby SuperNova factory.  We show that the
  image restoration obtained fulfills the stringent requirements on bias and
  photometricity needed by this program. The reconstruction yields sub-percent
  integrated residuals in all the synthetic filters considered both on real
  and simulated data.

  Even though our implementation is tied to the SNfactory data, the method
  translates to any hyper-spectral data. As such, it is of direct relevance to
  several new generation instruments like MUSE. Also, this technique
  could be applied to multi-band astronomical imaging for which image
  reconstruction is important, for example  to increase image resolution
  for weak lensing surveys.

\end{abstract}


%--------------------------------------------------------------- INTRODUCTION -
\section{Introduction}
Multi-channel data usually refers to images of the same scene observed at
multiple wavelengths, time frames, polarizations, etc. Data of this nature is
involved in a wide range of applications such as remote sensing, biological
and medical imaging, and of course astronomy. Because the light is spread out
on multiple channels instead of being integrated on a single image, the
information content of such data is increased at the cost of a lower signal to
noise or achievable resolution for the same exposure time. Extracting the
maximum of information from images in which each photon is acquired at a dear
cost is thus a recurring concern for multi-channel imaging. All the techniques
aiming toward this purpose, for example by removing the instrumental blur in
order to recover a higher quality image, fall under the denomination of image
restoration.

First attempts to restore multi-channel data made use of classical 2D
restoration techniques like Wiener filter or Richardson-Lucy algorithms on
each individual channel. The pitfall of these method is that they ignore the
natural cross-channel correlations existing in the data.  The first
restoration technique specifically dedicated to multichannel data was
introduced by \citet{Hunt1984}. It proposed a MMSE restoration filter (Wiener
filter) based on the assumption that the signal auto-correlation is spatially
and spectrally separable. This assumption was later relaxed by
\citet{Galatsanos1989}. Many other multichannel linear restoration filters
\citet{Katsaggelos1993} have been proposed since, using Kalman filters
\citep{Tekalp1990}, adaptive 3D Wiener filter \citep{Gaucel2006} or
regularized least square \citep{Galatsanos1991}. More recently
\citet{Benazza-Benyahia2006} and \citet{Duijster2009} adapted a
Fourier/Wavelet restoration technique ( ForWarD, \citet[][]{Neelamani2004}) to
multispectral data.

To enhance spatial resolution of multispectral data many authors
\citep{Hardie2004, Sroubek2006,Molina2007, Vega2009, Vega2010} have proposed
to merge spatial information contained in high resolution panchromatic images
with the spectral information of low resolution hyper-spectral images.  Based
on this approach, a technique for spectral resolution enhancement has been
proposed by \citet{Akgun2005} while \citet{Bobin2009} performs spatial
resolution enhancement. His technique relies on the strong assumption that the
scene observed is the linear combination of only a few materials with unknown
spectrum.

As illustrated by these examples, most work on restoration of hyper-spectral
images is done for remote sensing and color (RGB) images. Those methods can't
directly be applied to astronomical data because of its specific features like
large dynamic range and strong sharp features (for example narrow emission
lines). To the best of our knowledge, restoration techniques for
hyper-spectral astronomical images have been proposed for slit spectrography
data \citep{Courbin2000,Lucy2003} or data composed of slit spectrography scans
with Spitzer \citep{Rodet2008} but no such technique has been proposed for 3-D
spectroscopy. By 3-D spectroscopy, we mean here \ie $(\Dir,\Wavelength)$ data
observed simultaneously via an Integral Field Spectrograph (IFS) with
$\Dir=(\DirLetter_1,\DirLetter_2)$ the 2-D angular position along the slit and
$\Wavelength$ the wavelength.

Even though \citet{Schulz1993} and following work present techniques of blind
deconvolution of multi-frame data cubes, they don't readily apply to 3-D
spectroscopy. They are developed on sequences of short exposure images of
turbulence-degraded observation of still object that have their own
specificity compared to astronomical IFS data cubes.  Maybe closer to the
technique we propose here, \citet{Schultz1995} discuss a method based on
\emph{maximum a posteriori} (MAP) spatial regularization with Hubber function.
This approach is similar to ours in that they also propose a spatio-spectral
weighting of their regularization. Nevertheless, their work targets color
images very different from 3-D astronomical data.

Even among other astrophysical data, Integral Field Spectroscopy needs a
specific approach. For example, WMAP CMB observations are multi-spectral
images with 5 disjoint spectral bands, and display relatively smooth maps. On
the other hand, IFS like SAURON, SNIFS, FLAMES+GIRAFFE, KMOS, WIFES, and in the
future, MUSE observe simultaneously the spectra of whole regions of the sky,
yielding $(\Dir,\Wavelength)$ data cubes with several hundreds of wavelength
bins.  Each wavelength ``slice'' of such a cube is a mono-chromatic image at
the resolution of the IFU spectrograph, and can display peaked spatial
structures (like galactic cores and stars), as well as narrow emission or
absorption lines.  Dedicated image reconstruction techniques are needed to
take full advantage of the data gathered by these instruments.  The method
presented in this paper positions itself in this context: it is very generic
and exploits all the intrinsic continuities of hyper-spectral or data in order
to yield the best image reconstruction possible.

In section \ref{sec:model} we describe the hyper-spectral mathematical model
we wish to reconstruct. In section \ref{sec:inverse-problem} we present the
inverse problem approach we propose to achieve the hyper-spectral image
restoration. In section \ref{sec:image-reconstruction} we present the specific
context of the SuperNova factory in which our procedure is
exemplified. Section \ref{sec:summary} summarizes the algorithm
implemented. Sections \ref{sec:simul-result} and \ref{sec:data-result} show
the result of our procedure both on simulated and real data, and Section
\ref{sec:discussion} is dedicated to the discussion of the performance of the
algorithm and of possible future improvements.

\section{The Direct Model}
\label{sec:model}
In the following we describe how the model of the observed data relates to the
specific intensity  of the object of interest $\Iobj(\Dir,\Wavelength)$, with $\Dir =
(\DirLetter_1,\DirLetter_2)$ the 2-D position angle. 
The 3-D model of the observed data is a mixture of two components, the
observed object (the scene) and the background, convolved by the point spread
function (PSF) of the instrument. 
\begin{align}
  \label{eq:model_1}
  \Imodel(\Dir, \Wavelength) &= \iint \Bigl[\Iobj(\Dir', \Wavelength) +
  \Isky(\Wavelength)\Bigr] \, \PSF(\Dir - \Dir', \Wavelength)
  \, \D{\Dir'} 
\end{align}
with $\Isky(\Wavelength)$ the specific intensity of the sky (assumed to be
spatially uniform), $\PSF(\Dir,\Wavelength)$ the spatial point spread function
(PSF).
In order to simplify the equations, we will assume in what follows that we
work on post processed data, for which dark current and spectral calibration
have been properly handled.  

In \Eq{eq:model_1}, we assume that there is no cross-talk between spectral
channels, or that it has been properly accounted for at the CCD extraction
level. Formally, we do not work with CCD 2D images on which the pixel location
corresponds to a position and a wavelength, but with \emph{data cubes} where
the spectra have been properly extracted from the CCD and are labeled by their
2D position in the field of view. We also assume that the PSF is stationary
(shift-invariant) which is appropriate for small fields of view.  The PSF is
not necessarily normalized in order to account for the variable throughput
(atmospherical and instrumental transmission):
\begin{equation}
  \label{eq:throughput}
  \eta(\Wavelength) = \iint\PSF(\Dir,\Wavelength)\,\D{\Dir} \, .
\end{equation}
Similarly, the wavelength-wise PSF's may be centered at a location
$\Dir_{\Wavelength}$ which depends on the wavelength so as to account for
imperfect instrumental alignment and atmospherical differential refraction
(ADR).  Finally, to conserve the physical units of the quantities of interest, the
angular area of the pixel and the effective spectral bandwidth can also be
integrated in the PSF.

Introducing the specific intensity of all background sources:
\begin{align}
  \label{eq:background}
  \Ibg(\Wavelength) &= \Isky(\Wavelength) \,
  \iint\PSF(\Dir,\Wavelength)\,\D{\Dir}\\
  &= \eta(\Wavelength)\,\Isky(\Wavelength) 
\end{align}
our model equation simplifies to:
\begin{equation}
  \Imodel(\Dir, \Wavelength)
  = \iint \Iobj(\Dir', \Wavelength) \, \PSF(\Dir - \Dir', \Wavelength)
  \, \D{\Dir'} + \Ibg(\Wavelength) \, .
  \label{eq:model-2}
\end{equation}

After background subtraction, the available data is a \emph{cube} of
quasi-monochromatic images that we can model as follows:
\begin{equation}
  \label{eq:discrete-data}
  \Data[j,\ell] = \left[\Imodel(\Dir_j, \Wavelength[\ell]) - \Ibg(\Wavelength[\ell])\right]
  \,\Delta\DirLetter_j^2\,\Delta\Wavelength[\ell] 
  + \Error[j,\ell]
\end{equation}
with $\Dir_j$ the angular direction of $j$-th image pixel, $\Wavelength[\ell]$
the effective wavelength in $\ell$-th spectral channel, and $\Error[j,\ell]$ an
error term due to the noise and the model approximations. Multiplication by
$\Delta\DirLetter_j^2$, the area of $j$-th pixel, and by
$\Delta\Wavelength[\ell]$, the effective bandwidth of the $\ell$-th spectral
channel, approximately takes into account the integration of the signal.

Without loss of generality, the continuous 3-D spatio-spectral distribution of
the object we wish to reconstruct can be parametrized by a finite number of
coefficients by means of expansion onto a basis of functions, for instance:

\begin{equation}
  \label{eq:object-model}
  \Iobj(\Dir,\Wavelength) = \sum_{k,\ell'}
  \Param[k,\ell'] \, \SpatialSpline(\Dir - \GridDir[k])
  \, b^\Tag{spectral}_{\ell'}(\Wavelength)
\end{equation}
with $\SpatialGrid=\{\GridDir[k]\}_{k=1}^{\Grid{\Nfov}}$ the grid of sampled
angular positions in the synthesized field of view, $\SpatialSpline(\Dir)$ the
2-D angular interpolation (or basis) function,
$\SpectralGrid=\{\GridWavelength[\ell']\}_{\ell'=1}^{\Grid{\Nwave}}$ the grid
of sampled wavelengths and $\SpectralSpline(\Wavelength)$ the spectral
interpolation (or basis) function. $\Grid{\Nfov}$ is the number of
\emph{pixels} in the synthesized field of view and $\Grid{\Nwave}$ is the
number of spectral channels, or wavelength bins over which the object of
interest is described.  While the angular grid must be evenly spaced, this is
not the case for the spectral grid. This could be used by tuning the grid to a
finer resolution where sharp spectral features are expected. If the spectral grid is regular (as will be the case in our examples) the spectral interpolation function in \Eq{eq:object-model} becomes $b^\Tag{spectral}_{\ell'}(\Wavelength) = \SpectralSpline(\Wavelength - \GridWavelength[\ell'])$.
Note also that in order to avoid edge effects such as field aliasing, the spatial grid must 
be larger than the observed field of view. This point will be addressed in
more detail later on.

Using the parametrization of the object brightness distribution in
\Eq{eq:object-model} and the image model in \Eq{eq:model-2}, our
model of the data writes:
\begin{equation}
  \label{eq:discrete-model-explicit}
  \Data[j,\ell] = \sum_{k,\ell'} \LinOp[j,\ell,k,\ell']\,\Param[k,\ell'] + \Error[j,\ell]
\end{equation}
where the coefficients of the \emph{effective PSF} $\LinOp$ are:
\begin{align}
  \label{eq:psf-coefs}
  \LinOp[j,\ell,k,\ell'] & = \iint \SpatialSpline(\Dir - \GridDir[k]) \,
  \PSF(\Dir_j - \Dir, \Wavelength[\ell])
  \, \D{\Dir} \notag \\
  & \quad \times \SpectralSpline(\Wavelength[\ell] - \GridWavelength[\ell'])
  \,\Delta\DirLetter_j^2\,\Delta\Wavelength[\ell] \notag \\
  &= (\PSF\star\SpatialSpline)(\Dir_j - \GridDir[k])\,
  \SpectralSpline(\Wavelength[\ell] - \GridWavelength[\ell'])
  \,\Delta\DirLetter_j^2\,\Delta\Wavelength[\ell]
\end{align}
where $\star$ denotes 2-D convolution over the angular direction.  Using a
matrix notation, \Eq{eq:discrete-model-explicit} simplifies to:
\begin{equation}
  \label{eq:discrete-model}
  \Data = \LinOp\cdot\Param + \Error \, ,
\end{equation}
with $\Data$ the data vector, $\Error$ the noise vector, $\Param$ the
parameters describing the object of interest, and $\LinOp$ the linear operator
which approximates the convolution by the effective PSF and the sampling by
the detectors.

The angular and spectral step sizes can be chosen to match the effective
angular and spectral resolutions of the data to reduce the number of model
parameters. It can also be made finer in order to increase the resolution of
the reconstruction. We advocate to control the effective number of free
parameters by means of regularization. For the implementation case considered
in this paper, we chose to take the same angular grid as the detector pixels
and the same spectral grid as the effective wavelengths of the spectral
channels. In order to simplify the equations to come, we chose the cardinal
sine, $\mathrm{sinc}(\ldots)$, as the basis function. In summary, under these
prescriptions the model parameters simplify to the following discretization:
\begin{align}
  \label{eq:discrete-parameters}
  \Param[k,\ell] &= \Iobj(\Dir[k],\Wavelength[\ell])
\end{align}
where $\Wavelength[\ell]$ is the effective wavelength in the $\ell$-th
spectral channel of the data and $\Dir[k]$ is the $k$-th angular position in
an evenly spaced rectangular grid of \emph{pixels} which has the same sampling
step than the observed data. In the future, we could explore the use of
different basis like for example Gaussians as was done by \cite{Rodet2008} or
B-splines \citep{Thevenaz_et_al-2000-interpolation_revisited} in order to allow for the maximum of
the reconstruction to have an arbitrary location instead of being located at
the center of a pixel.

Note also that the  discretization presented here does not
account for convolution by CCD pixel although this can be built into the PSF
components. This would matter for blind deconvolution, but there is no
ambiguity here, where we suppose the PSF to be known. In addition, we remind
that in our model the PSF can account for all effects like differential
atmospheric refraction, throughput variations and pointing variations.

As we use the same spectral sampling in the restored cube $\Param$ than in the
data cube $\Data$, and since the cross talk was supposed either negligible or
accounted for at the CCD extraction step, the matrix $\LinOp $ is block
diagonal along the spectral dimension.  Formally,
\Eq{eq:discrete-model-explicit} and \Eq{eq:psf-coefs} become
\begin{equation}
  \label{eq:discrete-model-explicit-2}
  \Data[j,\ell] = \sum_{k} \LinOp[j,k,\ell]\,\Param[k,\ell] + \Error[j,\ell]
  \, ,
\end{equation}
with
\begin{align}
  \label{eq:psf-coefs-2}
  \LinOp[j,k,\ell]
  &= (\PSF\star\SpatialSpline)(\Dir_j - \GridDir[k])\,
  \,\Delta\DirLetter_j^2\,\Delta\Wavelength[\ell] \, .
\end{align}

With isoplanatic PSF, applying $\LinOp$ consists in $\Nwave$ discrete spatial
convolutions for each spectral channels.  Due to the convolution process, flux
from some part of the object just outside of the field of view is measured
inside data. To take correctly this fact into account, the estimated object
has to be spatially larger than the observed field of view. To that end, at
least half of the PSF support must be added on each side of the observed field
of view to form the restored field of view. Further, as in practice the
convolution is computed using FFT, the restored field of view has to be
extended even more in order to avoid edge artifacts. To have a correct
estimation of the object inside the restored field of view, one must work with
arrays of width (height resp.)  greater or equal to sum of the width (height
resp.) of restored and observed fields of view. Therefore, the application of
$\LinOp$ requires $\Nwave$ spatial FFTs of at least $\Grid{\Nfov} = 4\,\Nfov$
pixels, corresponding to about $4 \, \Nwave\,\Nfov\,\log_2(4\,\Nfov)$ complex
multiplications.

\section{Inverse Problem Approach}
\label{sec:inverse-problem}
The problem of restoring the parameters $\Param$ describing the object of
interest is a deconvolution problem that we chose to solve by an inverse
problem approach using the model of the data described above.
Deconvolution is a well known ill-posed problem which can be solved by adding
priors in a classical Maximum A Posteriori (MAP) or penalized likelihood
framework \citep{Bertero1998}. This is achieved by estimating the object
$\Param^+$ that minimizes the cost function $f(\Param)$:
\begin{equation}
\Param^+ = \argmin_{\Param}  f(\Param)\,,
\end{equation}
with 
\begin{equation}
  \label{eq:total-penalty}
  f(\Param)=\Fdata(\Param)+\Fprior(\Param)\,.
\end{equation}
This cost function $f(\Param)$ is the sum of a \emph{likelihood term}
$\Fdata(\Param) $ ensuring the agreement between the model $\Model $ and the
data $\Data$, and a \emph{regularization penalty} $\Fprior(\Param)$
introducing \emph{a priori} knowledge about the object.

\subsection{Likelihood and Noise Statistics}

Assuming Gaussian noise, the likelihood term reads:
\begin{equation}
  \label{eq:Fdata-Gen}
  \Fdata(\Param) = \Residual\T \cdot \Werror \cdot \Residual
\end{equation}
with the \emph{residuals}:
\begin{equation}
  \label{eq:residuals}
  \Residual = \Data - \LinOp\cdot\Param\,,
\end{equation}
where the weighting matrix $\Werror=\Cerror^{-1}$ is the inverse of the
spatio-spectral covariance of the noise and $\Param$ are the parameters.

For uncorrelated noise, $\Werror$ is diagonal and \Eq{eq:Fdata-Gen}
simplifies to:
\begin{equation}
  \label{eq:Fdata}
  \Fdata(\Param)=\sum_{j,\ell} w_{j,\ell}\,\ResidualLetter_{j,\ell}^{2}
\end{equation}
where the statistical weight $w_{j,\ell}$ is the reciprocal of the variance of
the errors at pixel $j$ and channel $\ell$.  This model can cope with
non-stationary noise and can be used to express confidence on measurements on
each pixel of the data. Since unmeasured data can be considered as having
infinite variance, we readily deal with missing or bad pixels as well as
spaxels outside of the observed field of view as follows:
\begin{equation}\label{eq:defW}
  w_{j,\ell} \bydef
  \left\{\begin{array}{ll}
  \Var(y_{j,\ell})^{-1} & \mbox{if $y_{j,\ell}$ is measured,}\\
  0 & \mbox{otherwise.}\\
  \end{array}\right.
\end{equation}
where a spaxel denotes the spatial delimitation of a bin, while a pixel
denotes the spatial and spectral delimitations of a bin.  This procedure
allows for proper accounting of bad pixels and  synthesized field of
view larger than the data support.

Except for very low detector noise (less than a few $e^-$ per pixel), we can
approximate the total noise (Gaussian detector noise plus Poissonian signal
noise) by a non stationary Gaussian noise \citep{Mugnier2004}:
\begin{equation}\label{eq:defW_poisson}
  w_{j,\ell} \bydef
  \left\{\begin{array}{ll}
  \left(\gamma   \max(y_{j,\ell},0)\,+\,\sigma^2_{j,\ell}\right)^{-1}
  & \mbox{if $y_{j,\ell}$ is measured,}\\
  0 & \mbox{otherwise,}\\
  \end{array}\right.
\end{equation}
where $\gamma$ accounts for the gain of the detector and
$\sigma^2_{j,\ell}$ is the variance of other approximately Gaussian noise (for
example read-out noise) on the data spaxel $(j,\ell)$.

\subsection{Regularization}
\label{sec:regul}

Most effective regularization methods account for the continuities along the
dimensions of the brightness distribution one wishes to reconstruct.  To that
end, it is customary to minimize the quadratic norm of finite differences.  In
the case of interest, our prior assumption is that the real distribution
should be rather smooth, which should thus also be the case of the
reconstructed 3-D image. Even for images with a peaked galaxy core this
assumption holds, since a noisy image will always be less smooth than the
original image we wish to reconstruct. Extending these considerations to all
the dimensions of interests, we chose a regularization term that writes:
\begin{align}
  \Fprior(\Param) &= \sum_{\V{k},\ell,\Delta{}\V{k}} w^\Tag{spatial}_{\V{k},\ell,\Delta{}\V{k}}\,
  \left(\Param_{\V{k}+\Delta{}\V{k},\ell} - \Param_{\V{k}, \ell}\right)^2 \notag\\
  &\quad + \sum_{\V{k},\ell,\Delta{}\ell} w^\Tag{spectral}_{\V{k},\ell,\Delta{}\ell}\,
  \left(\Param_{\V{k},\ell+\Delta{}\ell} - \Param_{\V{k}, \ell}\right)^2 
  \label{eq:our-regul}
\end{align}
where the sum over $\Delta{}\V{k}$ and $\Delta{}\ell$ takes into account a
given neighborhood of voxel $(\V{k},\ell)$. The weights
$w^\Tag{spatial}_{\V{k},\ell,\Delta{}\V{k}} \ge 0$ and
$w^\Tag{spectral}_{\V{k},\ell,\Delta{}\ell} \ge 0 $ are used to tune the local strength of the regularization. 

In a Bayesian framework this expression would have been obtained provided that
the finite differences follow independent Gaussian distributions. Under these
assumption, their covariance matrix is diagonal and its inversion leads to
write the weights as:
\begin{align}
  w^\Tag{spatial}_{\V{k},\ell,\Delta{}\V{k}} &=
  \zeta_\Tag{spatial}\,\expectation{\left(\Param_{\V{k}+\Delta{}\V{k},\ell} - \Param_{\V{k}, \ell}\right)^2 }^{-1} \label{eq:theoretical-spatial-regul-weight}\\   
  w^\Tag{spectral}_{\V{k},\ell,\Delta{}\ell} &=
  \zeta_\Tag{spectral}\,\expectation{\left(\Param_{\V{k},\ell+\Delta{}\ell} - \Param_{\V{k}, \ell}\right)^2 }^{-1} \label{eq:theoretical-spectral-regul-weight}  \,,
\end{align}
where $\expectation{\cdots\!\,}$ denotes the expectation.
We introduce the fudge factors $\zeta_\Tag{spatial}$ and
$\zeta_\Tag{spectral}$ to account for the fact that strictly speaking, the
finite differences can not be independent since they derive from a smaller set
of parameters ($\sim 3$ times smaller in 3D).

The theoretical regularization weights in
\Eq{eq:theoretical-spatial-regul-weight} and
\Eq{eq:theoretical-spectral-regul-weight} are generally unknown.  For simple
2-D image restoration (\eg deconvolution), it is customary to assume that the
weight of the spatial regularization does not depend on the position. In other
words, the statistics of the spatial fluctuations are expected to be
stationary. Extending this prescription, one could similarly assume that the
regularization weights do not depend on the wavelength. This would reduce the
problem to the tuning of only two regularization hyper-parameters: the weight
of the spatial regularization and that of the spectral regularization.
Nevertheless, owing to the high dynamical range of astronomical images, a
procedure based on average hyper-parameters could lead to over-regularization
of large features or under-regularization of small features. Rather, we
suggest that the regularization weights be at least chromatic and that they
scale with the mean brightness of the object at a given wavelength.

Formally, this amounts to applying a stationnary regularization to the
spectrally flattened object:
\begin{equation}
  x'_{\V{k},\ell} = x_{\V{k},\ell}/s_{\ell}
\end{equation}
with $s_{\ell}=\avg{x_{\V{k},\ell}}_{\V{k}}$ the spatially averaged object spectrum,
$\avg{\cdots\,\!}_k$ denotes averaging over pixel index $\V{k}$. To avoid dealing
with non-linear regularization, we estimate the mean object spectrum
directly from the data:
\begin{equation}
  s_{\ell} = \avg{y_{j,\ell}}_j/\eta_{\ell}
\end{equation}
with $\eta_{\ell}=\eta(\Wavelength[\ell])$ the effective throughput in
$\ell$-th spectral channel, \cf \Eq{eq:throughput}.  Note that this
approximation is justified because we do not attempt to perform spectral
deconvolution and because we use the same wavelengths in the sought
distribution and in the data.

Finally, since we do not want to artificially introduce anisotropies, we
require spatial and spectral dimensions to be \emph{a priori} uncorrelated and
the spatial correlation to be isotropic. Thus, any spatio-spectral correlation
or spatial anisotropy in the reconstructed cube will be due to real effects
present in the data, not to our assumptions.

Given all those prescriptions, we write the regularization term as:
\begin{align}
  \label{eq:discrete-empirical-regul}
  \Fprior(\Param) &= \mu_\Tag{spatial} \,
  \sum_{k_{1},k_{2},\ell} D_{k_{1},k_{2},\ell}^{\Tag{spatial}}(\Param)
  \notag \\
  & + \mu_\Tag{spectral} \,
  \sum_{k_{1},k_{2},\ell} D_{k_{1},k_{2},\ell}^{\Tag{spectral}}(\Param)
\end{align}
with indices $(k_{1},k_{2},\ell)$ corresponding respectively to
$(\DirLetter_1,\DirLetter_2,\Wavelength)$ and where, for instance:
\begin{align}
  D_{k_{1},k_{2},\ell}^{\Tag{spatial}}(\Param) &=
   \left(\frac{\Param[k_{1}+1,k_{2},\ell]
   - \Param[k_{1},k_{2},\ell]}{s_{\ell}}\right)^2  \notag\\   
  &\quad + \left(\frac{\Param[k_{1}+1,k_{2},\ell] 
     - \Param[k_{1},k_{2},\ell]}{s_{\ell}}\right)^2
  \label{eq:discrete-spatial-D-operator}
  \intertext{and}
  D_{k_{1},k_{2},\ell}^{\Tag{spectral}}(\Param) &=
  \left(\frac{\Param[k_{1},k_{2},\ell+1]}{s_{\ell+1}} -
  \frac{\Param[k_{1},k_{2},\ell]}{s_{\ell}}\right)^2 \,,
  \label{eq:discrete-spectral-D-operator-1}
  \intertext{or}
  D_{k_{1},k_{2},\ell}^{\Tag{spectral}}(\Param) &= 
  \left(\frac{\Param[k_{1},k_{2},\ell+1]
    - \Param[k_{1},k_{2},\ell]}{\frac{1}{2}\,(s_{\ell} + s_{\ell+1})}\right)^2
  \,. 
  \label{eq:discrete-spectral-D-operator-2}
\end{align}
The difference between the two approaches is that
\Eq{eq:discrete-spectral-D-operator-1} biasses toward the mean spectrum
shape while \Eq{eq:discrete-spectral-D-operator-2} biasses toward a flat
spectrum.

Compared to the regularization defined in \Eq{eq:our-regul}, we restrict the finite differences
to $\Delta{}\V{k} \in \{(1,0),(0,1)\}$ and $\Delta{}\ell = 1$ and use the weights:
\begin{align}
  w^\Tag{spatial}_{\V{k},\ell, \Delta{}\V{k}}
  &\approx \mu_\Tag{spatial}/s^2_{\ell} \notag\\
  w^\Tag{spectral}_{\V{k},\ell, \Delta{}\ell}
  &\approx \mu_\Tag{spectral}/s^2_{\ell}\,.\notag
\end{align}

\section{3D spectroscopic image reconstruction}
\label{sec:image-reconstruction}
The method described in the previous sections is very general, and could be
applied to any set of multi-wavelength data. Because of the large number of
wavelength bin collected simultaneously, it is especially well suited to
Integral Field Spectroscopy data, like what will be collected by MUSE, or in
the case of our example, what has been observed by the SuperNova factory.

The SNfactory uses a micro-lens integral field unit named the SuperNova
Integral Field Spectrograph \citep[SNIFS,][]{Aldering2002} to observe type Ia
supernovæ.  SNIFS is a fully integrated instrument optimized for automated
observation of point sources on a diffuse background over the full optical
window at moderate spectral resolution. SNIFS
is mounted on the south bent Cassegrain port of the University
of Hawaii 2.2-m telescope (UH 2.2-m Mauna Kea)
and is operated remotely.  It consists of a high-throughput
wide-band pure-lenslet integral field spectrograph \citep[IFS, ``\`a la
  TIGER,''][]{Bacon1995}, a multi-filter photometric channel to image the
field surrounding the IFS for atmospheric transmission monitoring simultaneous
with spectroscopy, and an acquisition/guiding channel.  The IFU possesses a
fully filled $6'' \times 6''$ spectroscopic field-of-view subdivided into a
grid of $15 \times 15$ spatial elements (spaxels), a dual-channel spectrograph
covering 3200--5200 \AA\ and 5100-11000 \AA\ simultaneously, and an internal
calibration unit (continuum and arc lamps). Each spaxel is $0.43'' \times
0.43''$, which critically samples the PSF since the average seeing is $\sim
1.1''$.  To each spaxel corresponds a spectrum, obtained by dispersing the
light going through each one of the micro-lenses. Each observation thus
contains $15 \times 15$ spectra, each one corresponding to a specific
$(\DirLetter_1,\DirLetter_2)$ position in the field. Because of this 3D nature
$(\DirLetter_1,\DirLetter_2,\Wavelength)$, each exposure is named a datacube.

Each supernova is observed on the order of 10 times over a $\sim 50$ day
periods, yielding $\Nexp$ datacubes.  For each epoch the signal of the
datacube is that of a supernova, it's host galaxy and the sky, convolved by
the atmospheric and optics response named the Point Spread Function (PSF).
The average dataset for one supernova includes final references for each
spectroscopic channel. A final reference is an exposure taken when the
supernova has faded away, and which contains only galactic and sky light. The
background, or sky contribution for each exposure is considered to be flat on
the $6'' \times 6''$ field of the spectrograph. Also, due to the small size of
the field, the point spread function can be considered as shift invariant.

The supernovæ and the sky have spectra that vary with time, while the
galaxy spectro-spatial shape is invariant with time. The position
of the supernova relative to  the galaxy is also constant with time.  The PSF
and ADR vary with time and are calibrated from outside of the spectrograph
using simultaneous exposures obtained on a wider field imager. 
Using field stars from the same channel and standard stars observed with the
spectrograph, the datacubes we consider are wavelength and flux calibrated as
well as telluric corrected following the procedure that will be described in a
forthcoming collaboration paper. 

Since the scientific goal of the SuperNova factory relies on the analysis of
uncontaminated supernova spectra, the removal of all host galaxy contamination
is mandatory. In slit spectroscopy, the background is removed by using
neighboring galaxy spectra on the slit direction. Nevertheless, the amount of
galaxy light injected in the spectrum by the PSF from galaxy points outside of
the slit is very difficult (if not impossible) to constraint to the accuracy
needed by the SuperNova factory science goals. On the other hand, using the 3D
nature of the data collected with SNIFS, we can reconstruct the galaxy 3D
image, and given the PSF for each supernova exposure, subtract its
contribution inside of the field of view including the impact of galactic
structure from outside of the field. As the Nearby Supernova Factory
scientific goal relies crucially on the quality of the photometry of the SNe
Ia spectra, special care has been taken in order to reconstruct the galaxy
with minimal bias.

\section{Algorithm summary}
\label{sec:summary}

As discussed in \ref{sec:model} the specific characteristics of the galaxy
reconstruction grid are that in order to allow for field extrapolation, it has
to be larger than the data grid. Moreover, in order to accommodate the needs
of FFT convolution, we use a grid twice as large as that of the data. Also,
since for the SNfactory application we aim at a good photometric subtraction
of the host galaxy, but not (for now) at increasing the spatial nor spectral
resolution of the reconstructed image, we keep the same size of wavelength and
spatial bins in the reconstruction and in the data.  This choice of grid
allows for fast calculations, even though the formalism allows for increasing
the spectral or spatial resolution at the cost of an increase in computing
time

Based on the model described in section \ref{sec:model}, we implement the
following algorithm in order to reconstruct the galaxy datacube that will then
be used for host galaxy subtraction:

\begin{itemize}
\item \textbf{Read in all the calibrated ingredients}:
  \begin{itemize}
    \item The flux calibrated 3D final reference.
    \item The PSF grid: In order to allow for easy convolution, and in order
      to correctly account for the large scale impact of the PSF, the PSF grid
      is of the same size than the galaxy reconstruction. 
  \end{itemize}
\item \textbf{Estimation and subtraction of the spatially flat component of
  the data}.  Strong spectral features decrease locally the strength of our
  spatial regularization --- \Eq{eq:discrete-spatial-D-operator},
  \Eq{eq:discrete-spectral-D-operator-1} and
  \Eq{eq:discrete-spectral-D-operator-2} --- thus allowing for a larger
  granularity of the reconstruction around their peak wavelength.  Such
  behavior is highly undesirable when those spectral features are due to a
  spatially flat source. It is thus important to subtract even a crude
  estimation of the flat component of the data prior to the any of the
  reconstruction stages, including the computation of the regularization
  weights.  In practice, we estimate it using a classical iterative sigma
  clipping method.  Moreover, in a non negligible fraction of our data
  galaxies don't dominate over the entire spectrograph field. In this case,
  sky is the main contributor to the flat component we subtract. Its
  subtraction thus removes a large fraction of the background contamination
  from the galactic reconstruction that can then be better used for ancillary
  science.

\item \textbf{Estimation of $\Param^+$}: The galaxy reconstruction $\Param^+$
  is the solution of the minimization of the regularized cost function
  $f(\Param)$ presented in section \ref{sec:inverse-problem}. To solve this
  constrained minimization problem involving a large number of parameters
  ($>1\times10^5$) we use the VMLM-B algorithm \citep{Thiebaut2002} which is a
  limited memory variant of the variable metric method with BFGS updates
  \citep{Nocedal1999}. This algorithm has proved its effectiveness for image
  reconstruction and only requires the computation of the cost function
  together with its gradient. The memory requirement is a few times the size
  of the problem.  The hyper-parameters $\mu_\Tag{spatial} $ and
  $\mu_\Tag{spectral} $ are estimated by trial and error.  We first start
  tuning the two hyper-parameters alternatively, aiming for a value of
  $\Fdata(\Param)/N_{\textrm{data}} \sim 1$. To that end, we found it helpful
  to use the histograms of the residuals normalized by the noise standard
  deviation as displayed in fig.\ref{fig:fig4} . When their RMS is larger than
  1, it diagnoses an over-regularization, while values smaller than one at
  this step are usually symptomatic of an under-regularization. Once
  acceptable values are found, we resort to visual inspection of the
  residuals, where over smoothing of narrow lines or spatially localized
  structures in the residuals track respectively spectral or spatial
  over-regularizations. Conversely, large noise propagation to the spectrum or
  to the spatial map correspond to under-regularizations. Note that the best
  hyper-parameters are found for RMS of the above histograms a few percent
  below 1. We checked that once that we find a set of hyper-parameters to our
  liking, the results are stable on a neighborhood of these values.  In
  practice, thanks to our spectral flattening, when coping with real data from
  the SuperNova factory, we observe that the good values of these
  hyper-parameters are almost constant from one observation to another. The
  values used for SN2004gc presented in the following sections are
  $\mu_\Tag{spatial}=10^{-3}$ and $\mu_\Tag{spectral}=5\,10^{-3}$.
\end{itemize}
 
\section{Results on simulated data}
\label{sec:simul-result}
\subsection{Simulator}
In order to test the quality of our algorithm, and in an attempt to qualify
and quantify its 3-D image restoration capabilities, we implemented a
simulator that generates realistic SNfactory exposures. These exposures
contain a sky background, a galaxy and  possibly a supernova, with all these
components summed and convolved by a given PSF.

The sky spectrum is generated randomly from a PCA decomposition of all the
SNfactory sky backgrounds extracted from hostless SNe Ia exposures. This
procedure provides a sky that varies from one exposure to the other, while
displaying all the characteristic features of a real sky spectrum.

The galaxy is simulated from very finely spatially sampled models. Those
models are obtained by fitting a stellar population library
\citep{bruzual_stellar_2003} to real multi-color images
\citep{frei_catalog_1996}. The galaxy models provided this way display
realistic spectra as well as realistic spatial variations. Nevertheless, this
simulator lacks for now the ability to realistically include emission lines.

If need be, a supernova can also be added, as a point source atop of the
galaxy, following any spectral evolution template that one chooses to use. In
our specific case, since we are interested in checking the quality of the
galactic subtraction, we only simulated galaxy exposures, not including any
supernova. Moreover, in order to facilitate the comparison between the
reconstructed galaxy and the ground truth, we omit the sky.

Once the finely sampled model is built, including the galaxy and potentially
some sky background and a supernova, it is convolved by a finely sampled
PSF. It is then spatially integrated over the coarser grid of the
spectrograph.  The PSF used come from very finely sampled PSF models,
extracted from the photometric channel of SNIFS. This insures that the PSF
shape are realistic. In particular, it allows for non symmetric PSF shapes.

Finally, photon noise (with Poisson statistic) and read out noise (with
Gaussian statistic) are added. The photon noise level is selected by the user,
and depends on a target signal to noise ratio. The read out noise on the other
hand is fixed and reproduces the statistic of SNIFS read out noise.

This procedure yields simulated SNIFS exposures that are then reconstructed by
our algorithm following the exact same pipeline as used on real data.

\subsection{Results}

\begin{figure*}
  \centering
  \includegraphics[scale=0.4 ]{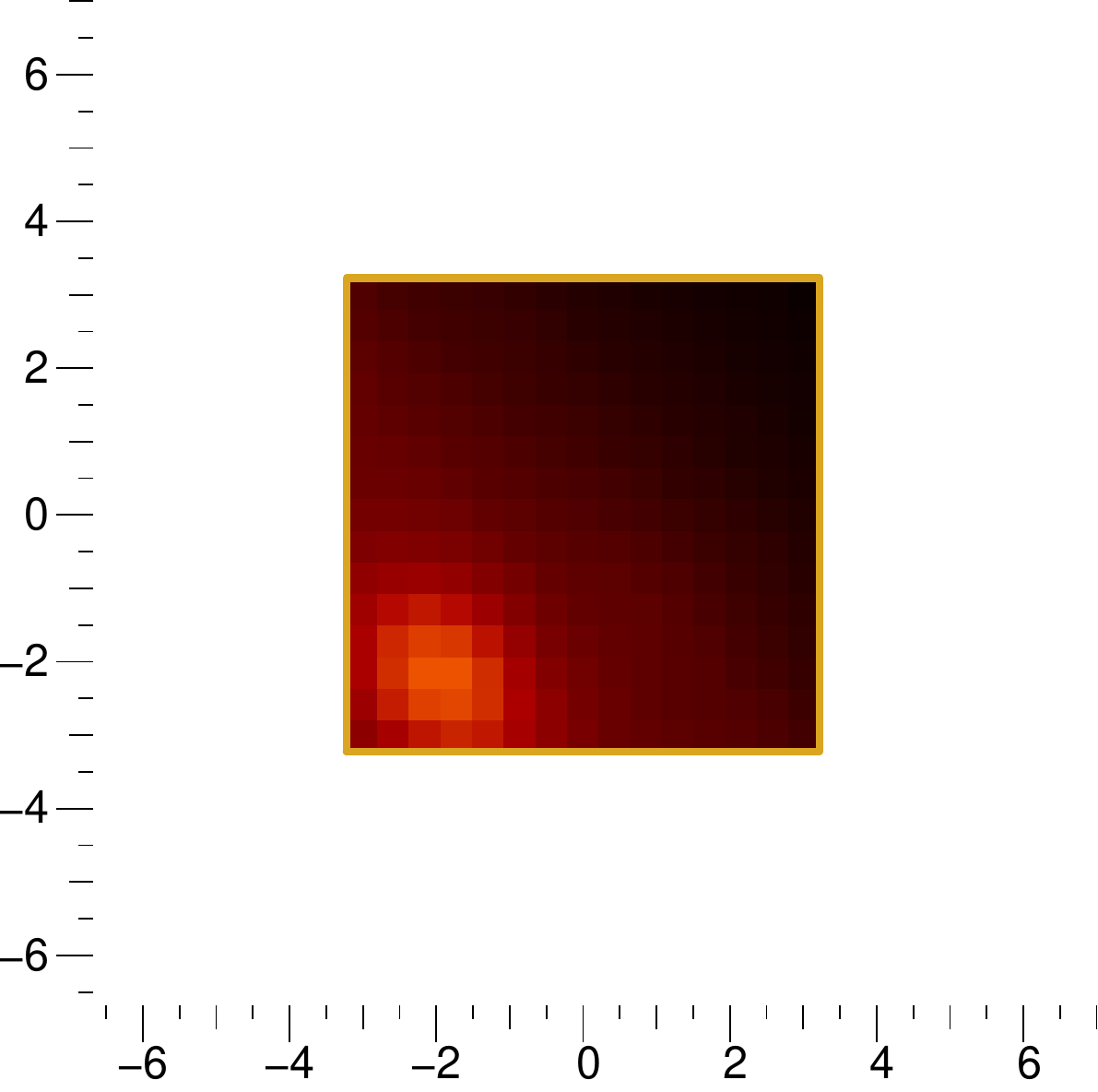}
  \hspace{1.5mm}
  \includegraphics[scale=0.4 ]{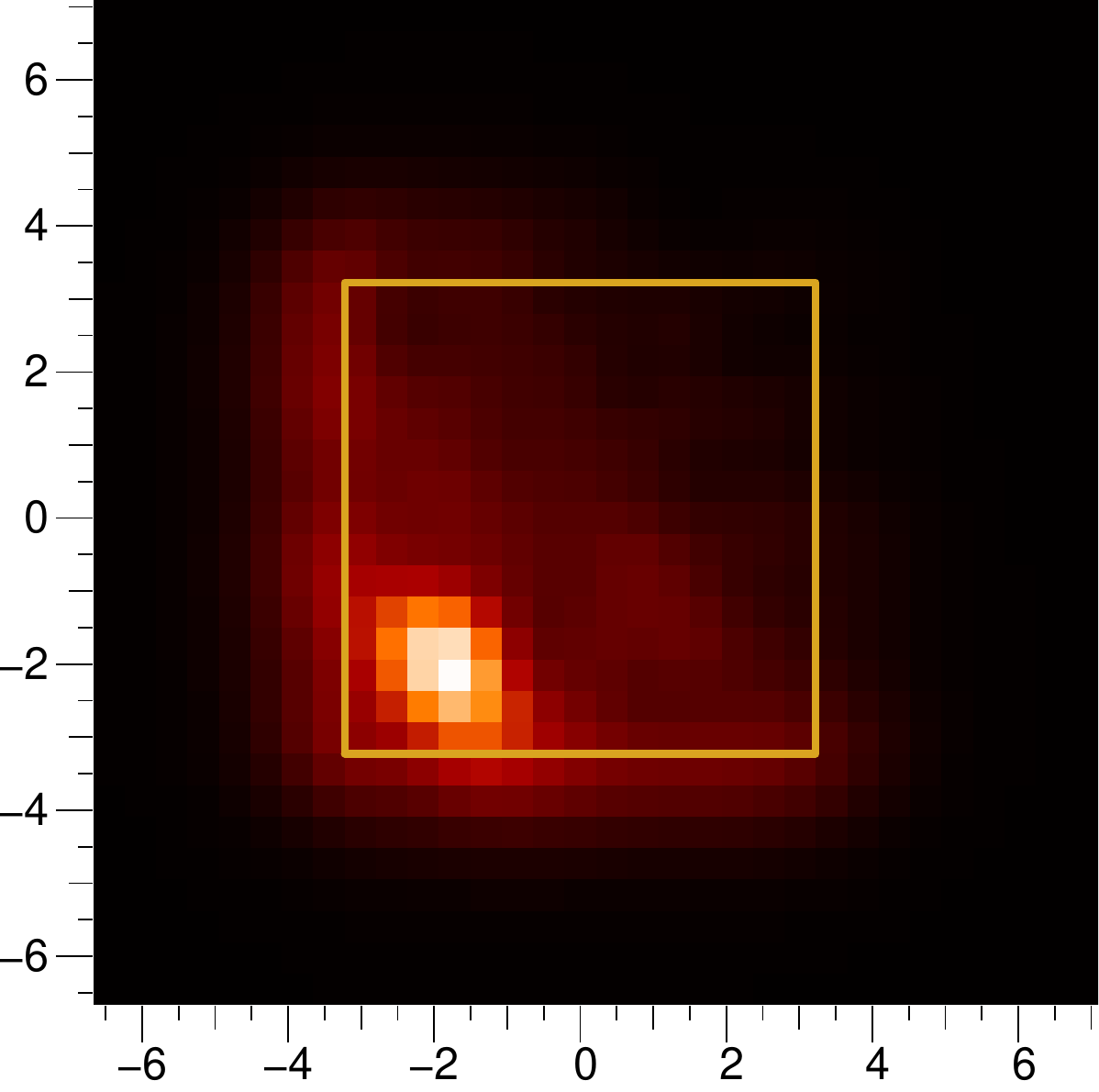}
  \hspace{1.5mm}
  \includegraphics[scale=0.4 ]{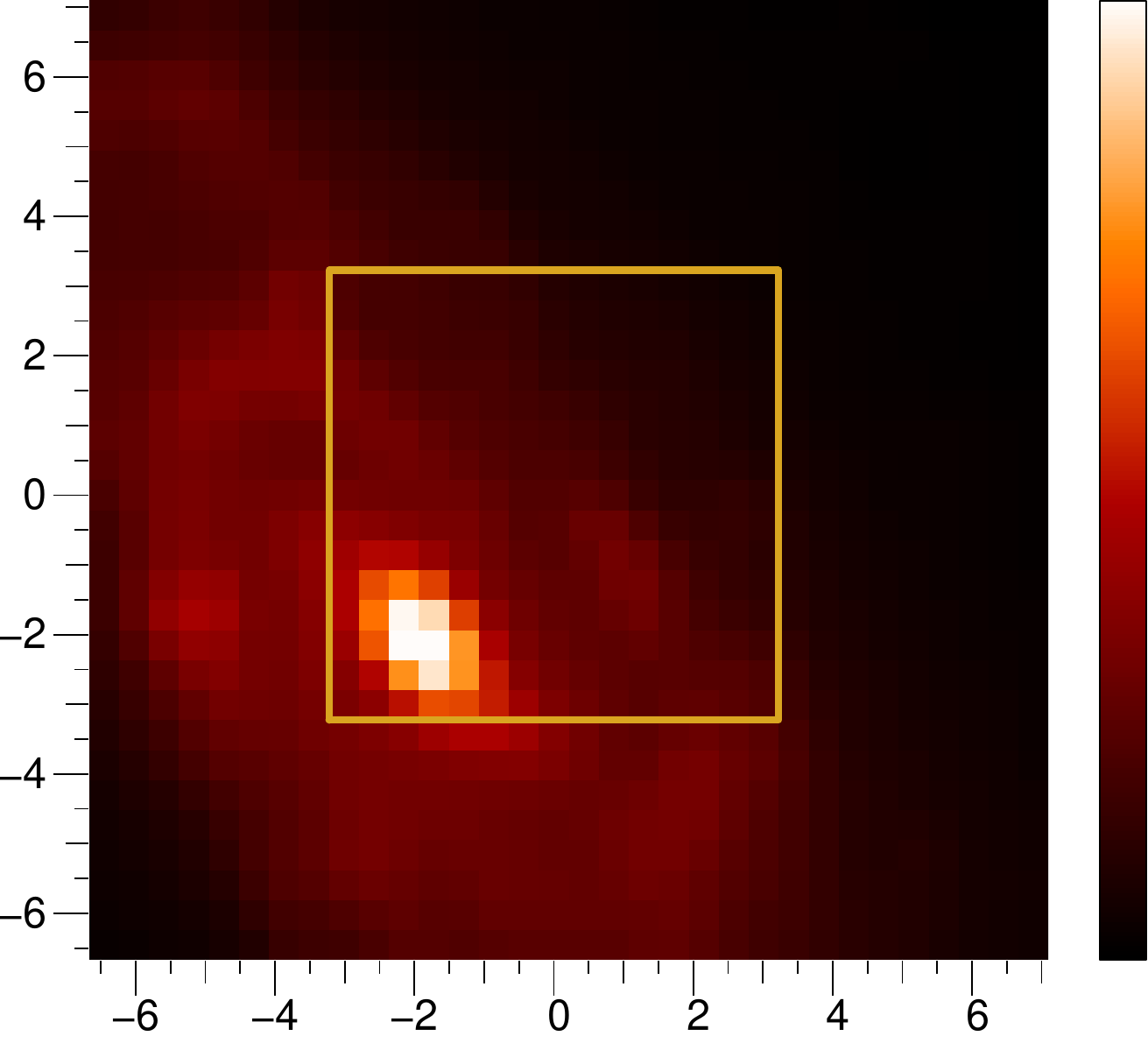}
  \caption{From left to right: data, reconstruction and ground truth,
    integrated over the SNf B band synthetic filter (top hat filter with a
    bandpass between 4102 \AA\ and 5100 \AA ). The color bar is the same for
    all 3 figures, and is in relative units between the minimum and the
    maximum of the ground truth signal image. The golden box in the
    reconstruction and the ground truth images represent the data support
    region.}
  \label{fig:fig1}
\end{figure*}

The datacube we use to test our reconstruction algorithm is a typical
SuperNova factory blue cube of $15\times15$ spaxels over $754$ wavelength bins
with an average signal to noise per spaxel and per wavelength bin of $\sim
8$. Figure \ref{fig:fig1} displays from left to right the data, the galactic
reconstruction and the ground truth integrated over wavelengths between
4102\,\AA\ and 5100\,\AA, which corresponds to a B band top-hat synthetic
filter. All three plots are displayed in the same color scale, showing that
the galactic core of the reconstruction is much more peaked than the one
observed in the data. Even after integration over $410$ wavelength bins,
almost no structure is discernible in the data. On the other hand, the
galactic arm is clearly visible in the reconstruction. Moreover, the galactic
arm on the top left and of the lower right of SNIFS field of view are well
extrapolated up to few spaxels outside of the field. The difference of Peak
Signal to Noise Ratio (PSNR) of the data and the reconstruction writes:
\begin{equation}
  \Delta\!\Tag{PSNR} = 10\,\log\left(
    \frac{\avg{(\Data - \Param_{\Tag{true}})^{2}}}
         {\avg{(\Param^{+} - \Param_{\Tag{true}})^{2}}}\right) \,,
\end{equation}
where $\avg{\ldots\!\,}$ denotes the average taken over all spaxels and
wavelengths. On the wavelength region corresponding to the top-hat B
filter used in 
Figure~\ref{fig:fig1}, we find an improvement of  $\Delta\!\Tag{PSNR}  = 5.2 \, \mathrm{dB}$.

\begin{figure*}
  \centering
  \includegraphics[scale=0.3]{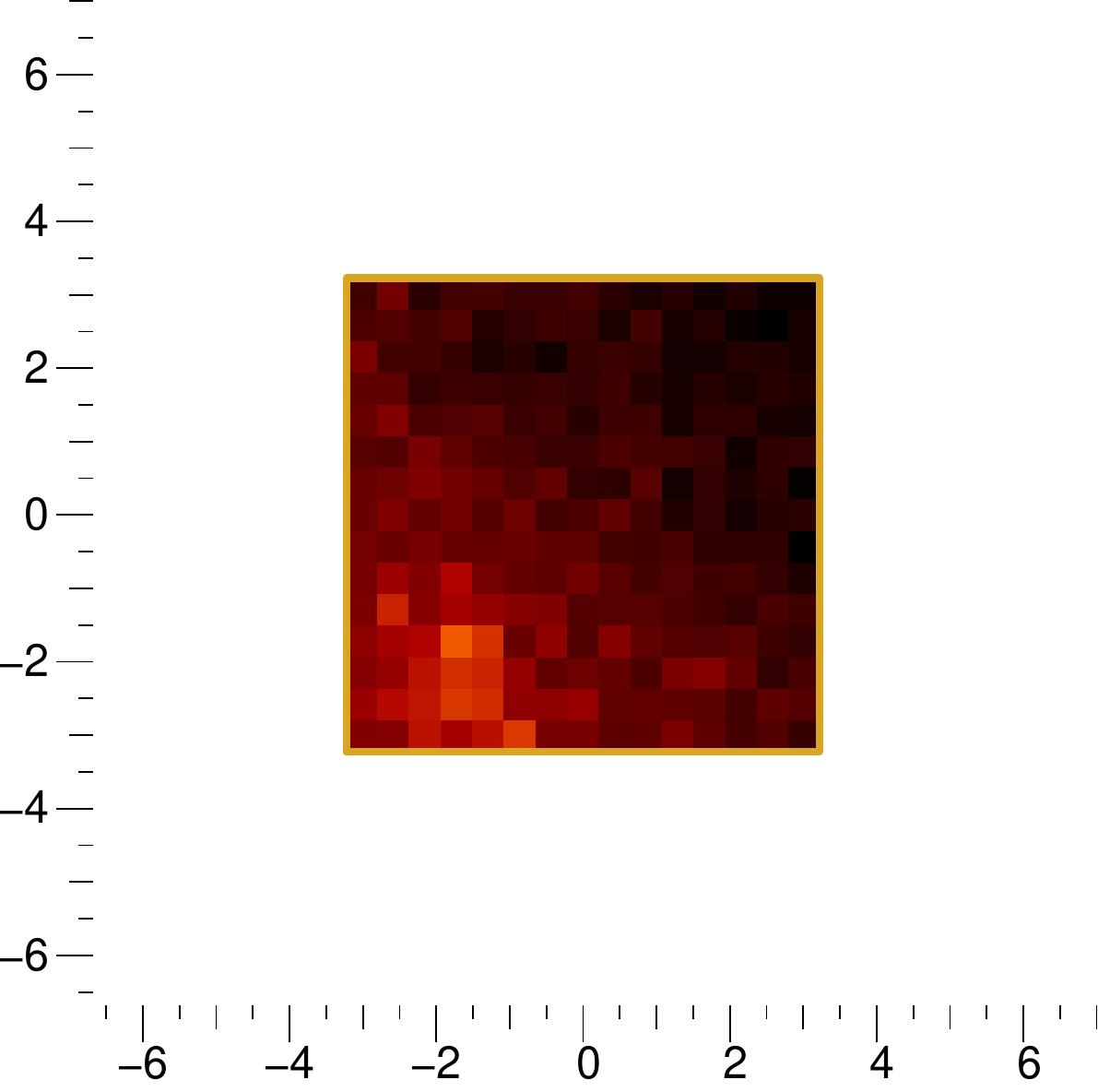}
  \hspace{1mm}
  \includegraphics[scale=0.3]{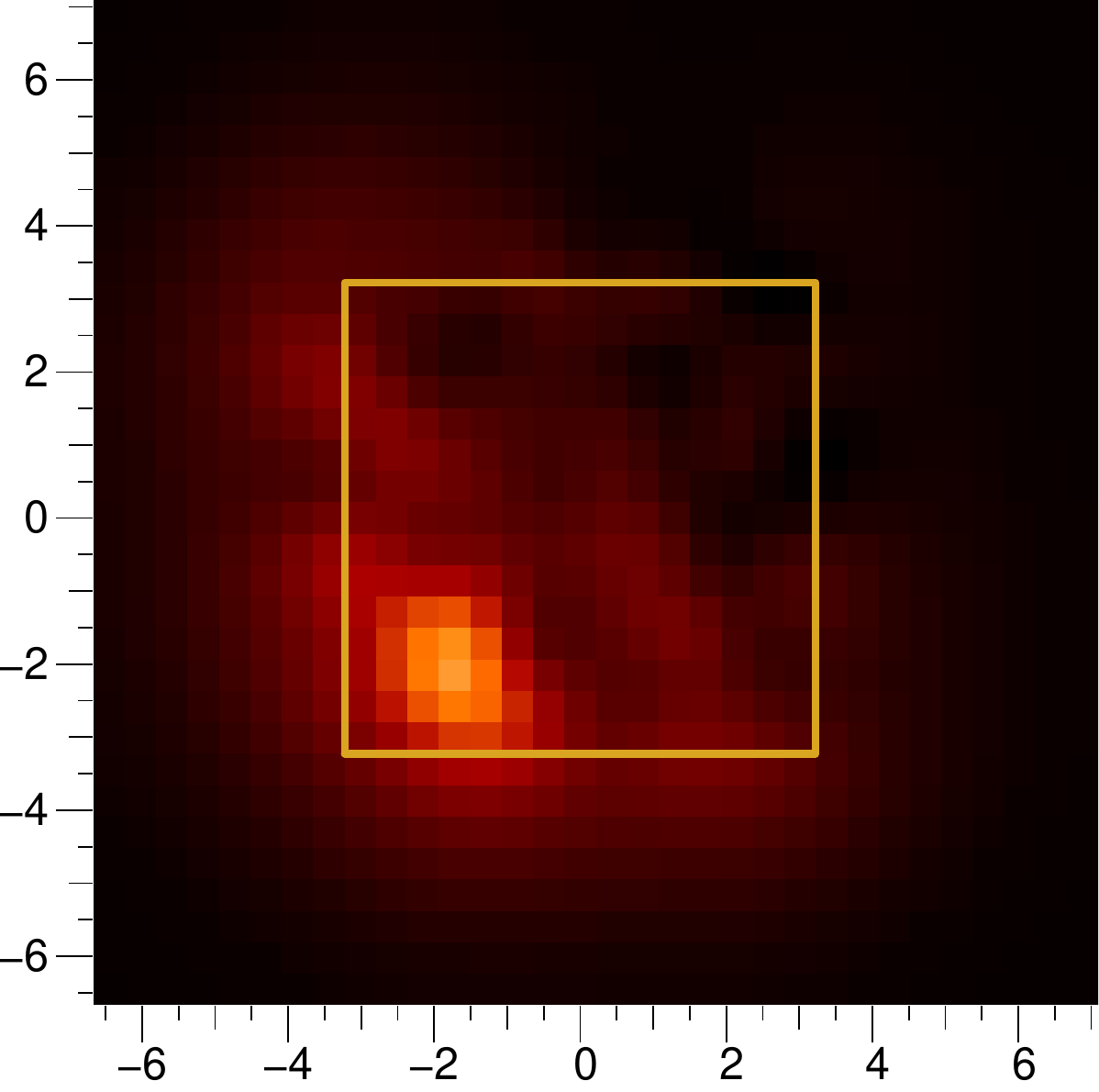}
  \hspace{1mm}
  \includegraphics[scale=0.3]{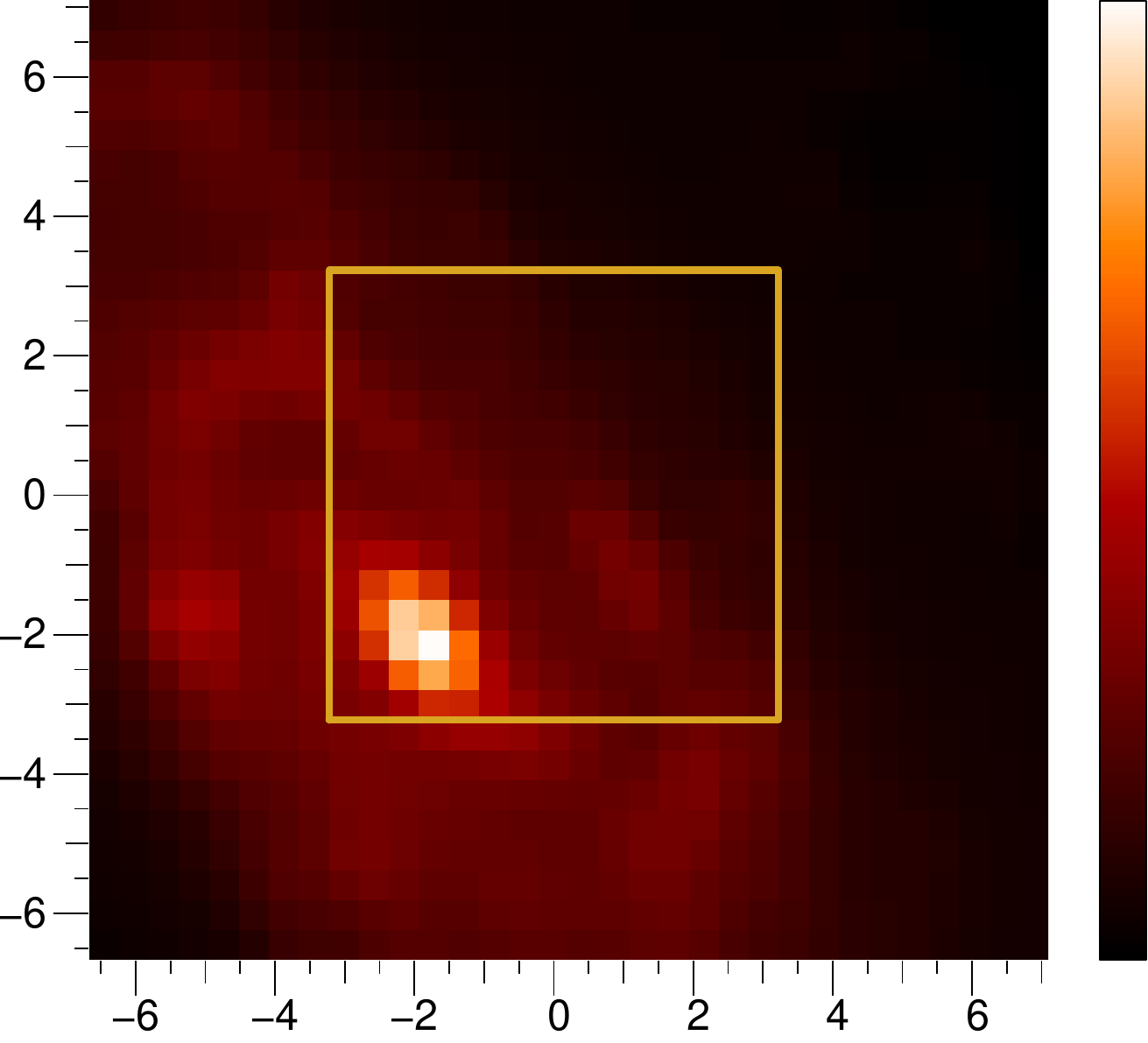}
  \caption{From left to right: data, reconstruction and ground truth for one
    single wavelength bin ($\lambda = 3654$\AA). The color bar is the same for
    all 3 figures, and is in relative units between the minimum and the
    maximum of the ground truth signal image. The golden box in the
    reconstruction and the ground truth images represent the data support
    region.}
  \label{fig:fig2}
\end{figure*}

A single wavelength version of fig.\ref{fig:fig1} is shown in
fig.\ref{fig:fig2}, for $\Wavelength=3654$\AA. The leftmost plot, which displays
the data slice, shows that at this wavelength the signal to noise is very
weak, as is the case for all wavelengths slices below the Balmer Break.
Nevertheless, the comparison between the galaxy reconstruction and the ground
truth show that for this slice the structure of the galaxy inside of the field
is well reproduced. The field extrapolation also catches the large upper left
galactic arm, and to some extent the lower right one. We observe that as
expected the extension of the field extrapolation is similar to the PSF FWHM
of $\sim 2.7$ spaxels or $\sim 1.1''$.

\begin{figure}
  \centering
  \fbox{
  \includegraphics[width=0.4 \textwidth]{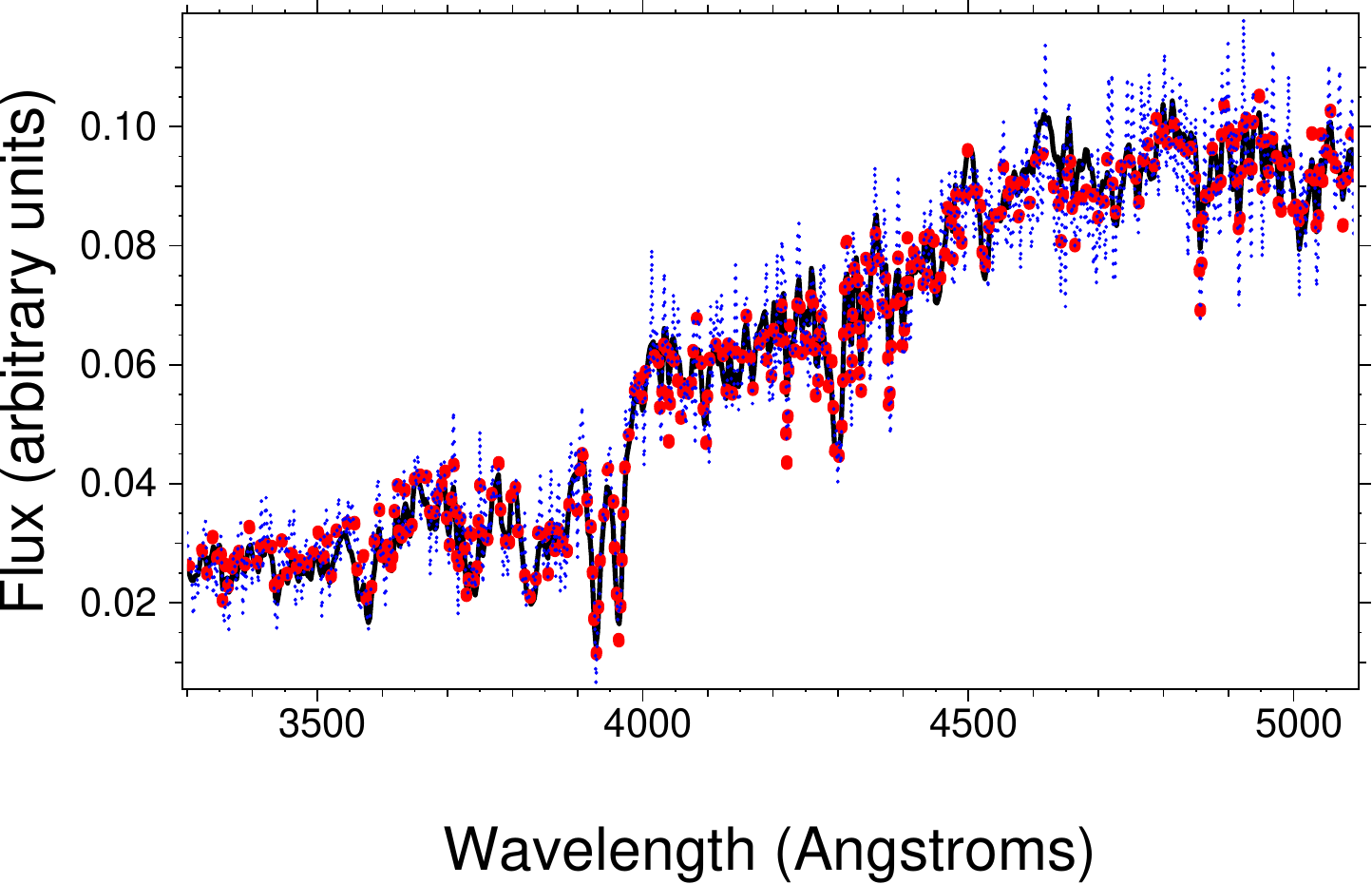}
  }
  \vskip 0.1cm
  \fbox{
  \includegraphics[width=0.4 \textwidth]{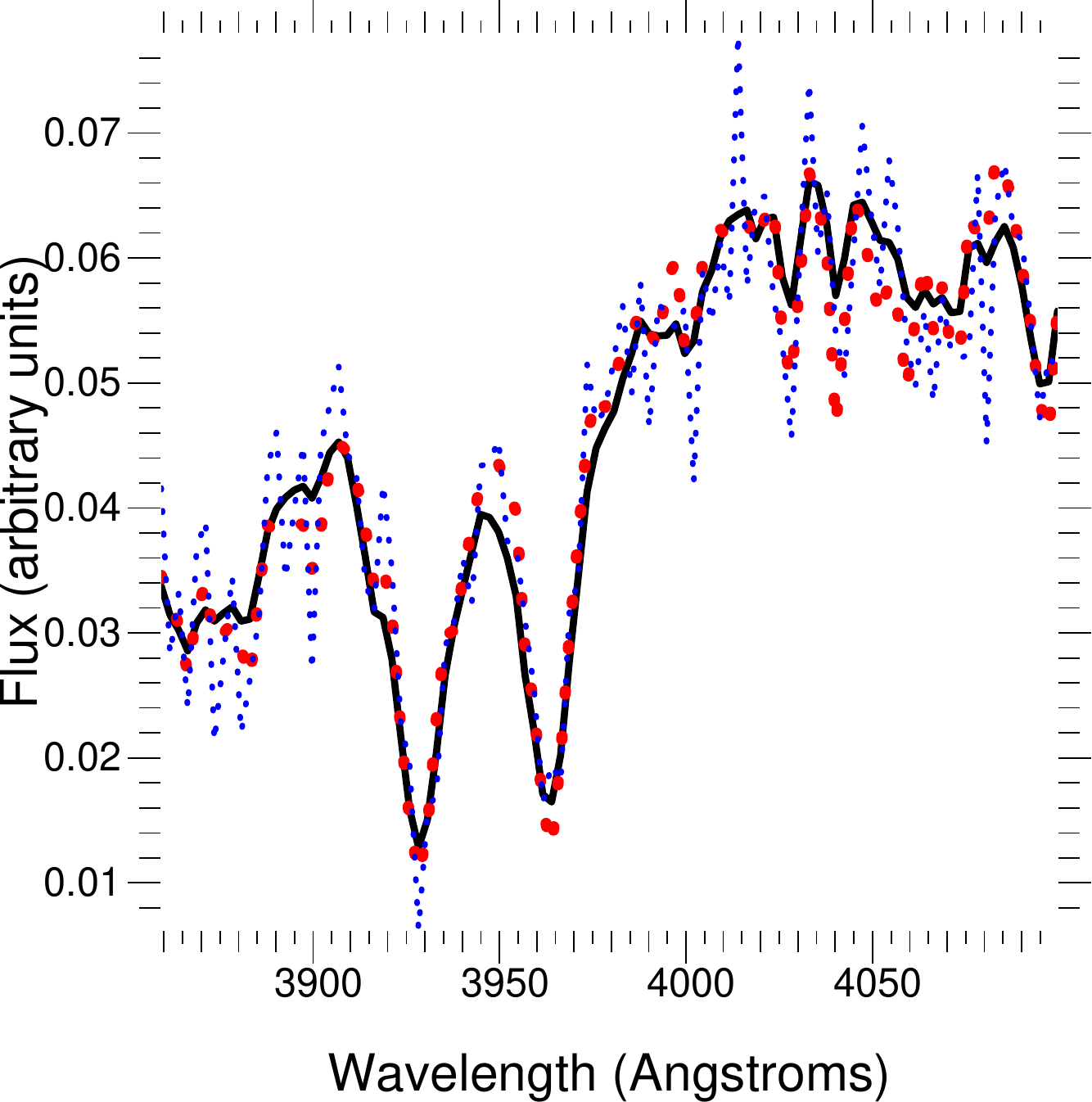}
  }
  \caption{Spectrum of the brightest spaxel of the reconstruction in black,
    superimposed over the same spaxel of the ground truth in red. Bottom plot
    is a zoom of the top plot on the region with largest spectral
    structure. In order to give a sense of the noise level of each spectra, we
    display in dotted blue the spectrum of the simulated data corresponding to
    the same location. Note also that in order to probe the quality of the
    relative reconstruction, and in order to remove the absolute differences
    due to variations of effective spatial resolution between the data, the
    truth and the reconstruction, we normalize all the spectra to the same
    average value}
  \label{fig:fig3}
\end{figure}

On the spectral side, Figure \ref{fig:fig3} displays the spectrum of the
brightest galaxy spaxel for the ground truth in red and for the reconstruction
in black. As stated above, we see that the galactic flux below 4000 \AA\ is
much lower than above. Nevertheless the galactic spectrum is well
reconstructed at all wavelengths. The lower panel of the plot shows a zoom
into the Ca H\&K strong features that are reproduced with minimal bias.
The spectrum of the data, displayed in dotted blue in both panels,
illustrates the fact that the regularization only biases the spectrum
reconstruction within the noise limits. For example the absorption trough at
$\sim 4300$\AA, larger than the Ca H\&K lines in the lower panel of the
figure, is reconstructed without bias. 

It has to be noted that if the spectral regularization slightly biases the
reconstructed spectrum, it also allows for a consistent and effective
reconstruction of the galaxy over the full wavelength range. Without such a
regularization, each wavelength slice would be considered independently, and
the noise level visible in the leftmost panel of Figure \ref{fig:fig2} would
make it impossible to reconstruct the galaxy with the accuracy displayed in
the middle panel of the same figure. The comparative gain of spectral
regularization and spectral flattening will be addressed in more details in
the next subsection (sec.\ref{sec:regul-discussion}).

\begin{figure}
  \centering
  \fbox{
  \includegraphics[width=0.45 \textwidth]{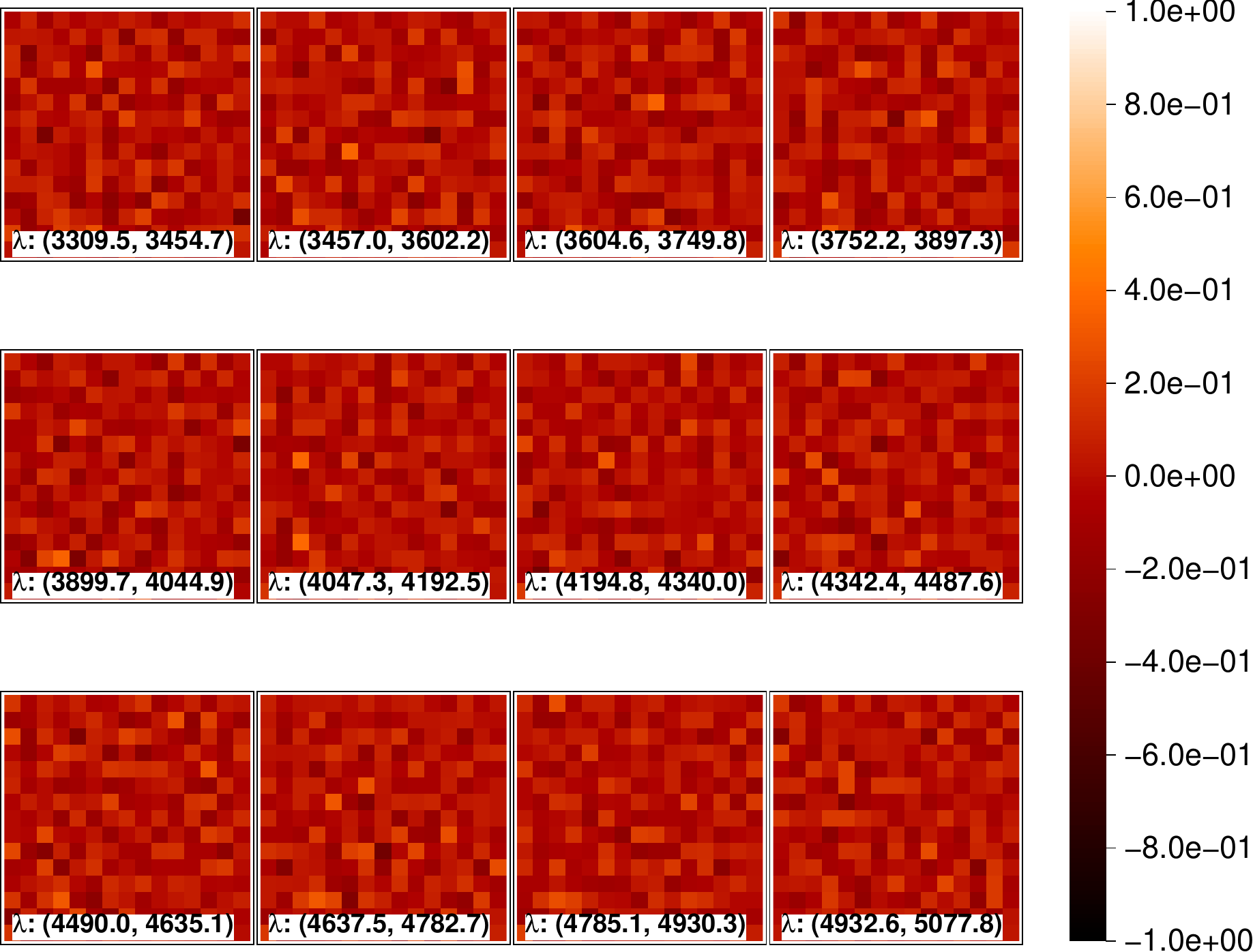}
  }
  \caption{Residual maps of the galaxy reconstruction over 12 consecutive
    wavelength slices (spectral integrals on $\sim 150$\AA\ wavelength
    bins). The residuals are normalized by the noise standard deviation on
    each pixel, giving a 2D vision of the histograms displayed in Figure
    \ref{fig:fig5} }
  \label{fig:fig4}
\end{figure}

\begin{figure}
  \centering
  \fbox{
  \includegraphics[width=0.45 \textwidth]{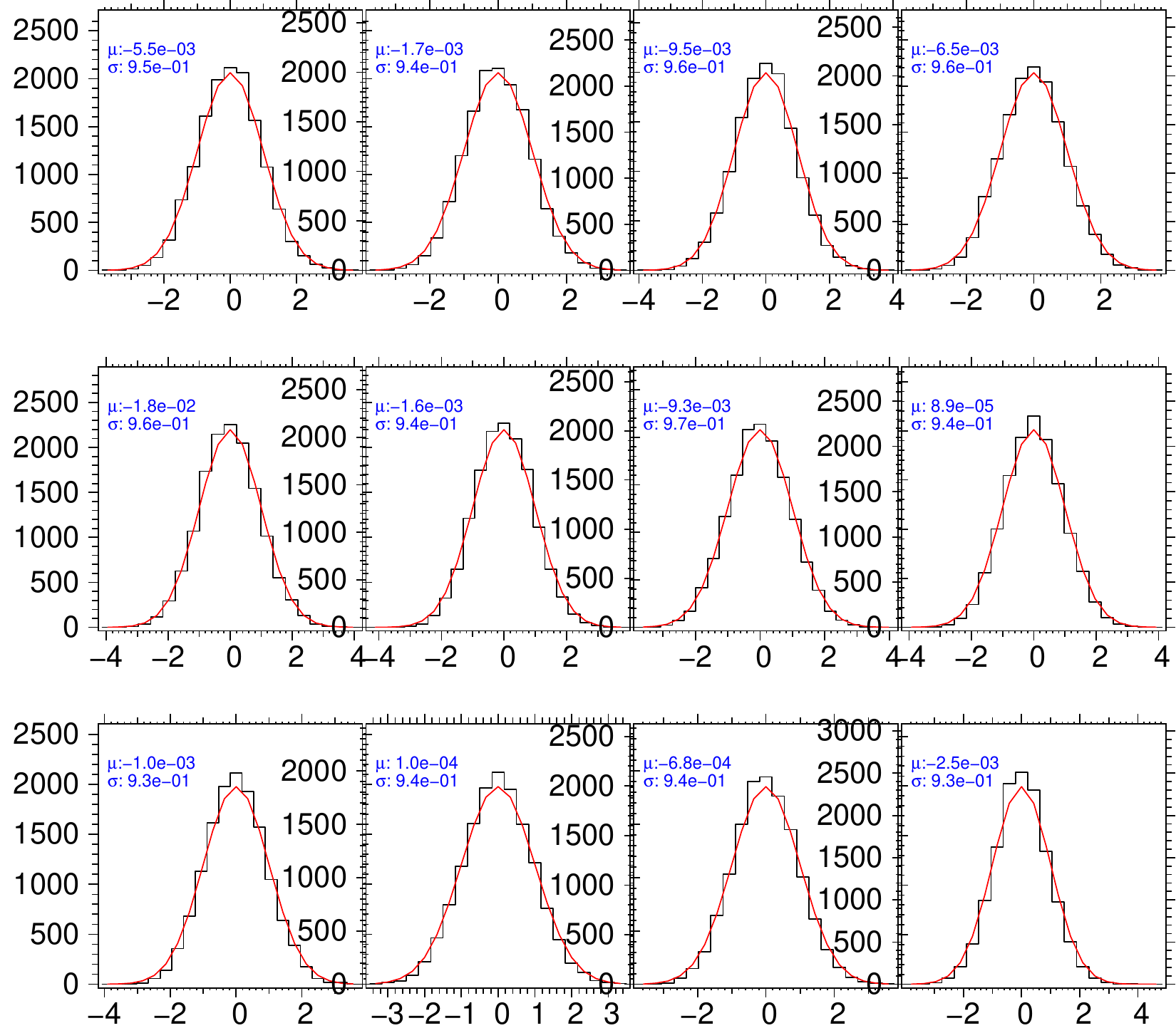}
  }
  \caption{Pull distribution of the residual slices displayed in
    fig.\ref{fig:fig2}. For each slice a Gauss curve of mean 0 and $\sigma=1$
    is plotted in red. The mean and standard deviation of the distribution of
    each slice are reported in blue.}
  \label{fig:fig5}
\end{figure}

Figure \ref{fig:fig4} presents the residual map comparing the data
and the model obtained once the reconstruction achieved. Its
twelve thumbnails integrate the residual over $\sim 150$\AA\ each
($\sim50$ wavelength bins), and the residuals are normalized by the error.
Figure \ref{fig:fig5} shows the histogram of these residuals normalized by the
error (also called pull distribution).  The mean values of the histograms,
that measure the reconstruction bias, are below the 1\% level for all but the
slice including the two deep Ca H\&K lines. For this later slice, the depth of
the Ca H\&K lines results in several wavelength bins with very low signal to
noise. Because we don't include the sky in this simulation, the absolute level
of the signal in the Ca H\&K trough is very small, leading to a small absolute
value of the variance. In turn, this small value of the variance used to
normalize the residual, results in a relatively large bias, even though its
absolute value is very small. The RMS values, close to 1, show that the
residuals are compatible with Gaussian noise. The systematic offset of
5\%, between the RMS values observed and 1 tends to diagnose a slight
over-fitting during the reconstruction process. This in turn will lead to
some noise to be included in the reconstruction. Nevertheless the other
results displayed previously show that this does not impact visibly on the
quality of the reconstruction. In practice, it means that reaching minimal
bias and an RMS within few percents of 1 is a good convergence criterion for
trial and error estimation of the hyper-parameters.

\subsection{Comparison between regularization schemes}
\label{sec:regul-discussion}

\begin{figure*}
  \centering
  \includegraphics[scale=0.3]{fig_2b.pdf}
  \hspace{1mm}
  \includegraphics[scale=0.3]{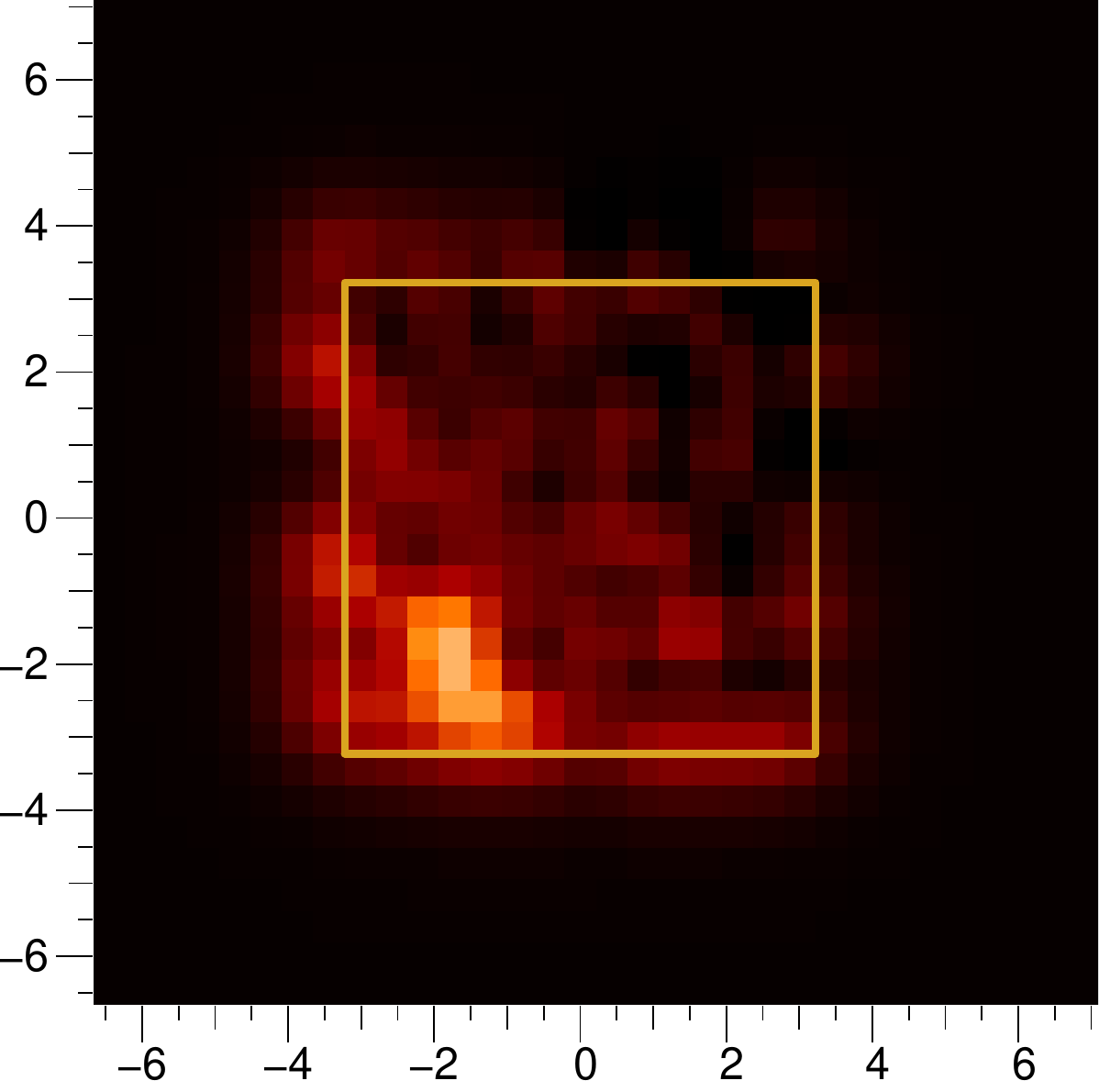}
  \hspace{1mm}
  \includegraphics[scale=0.3]{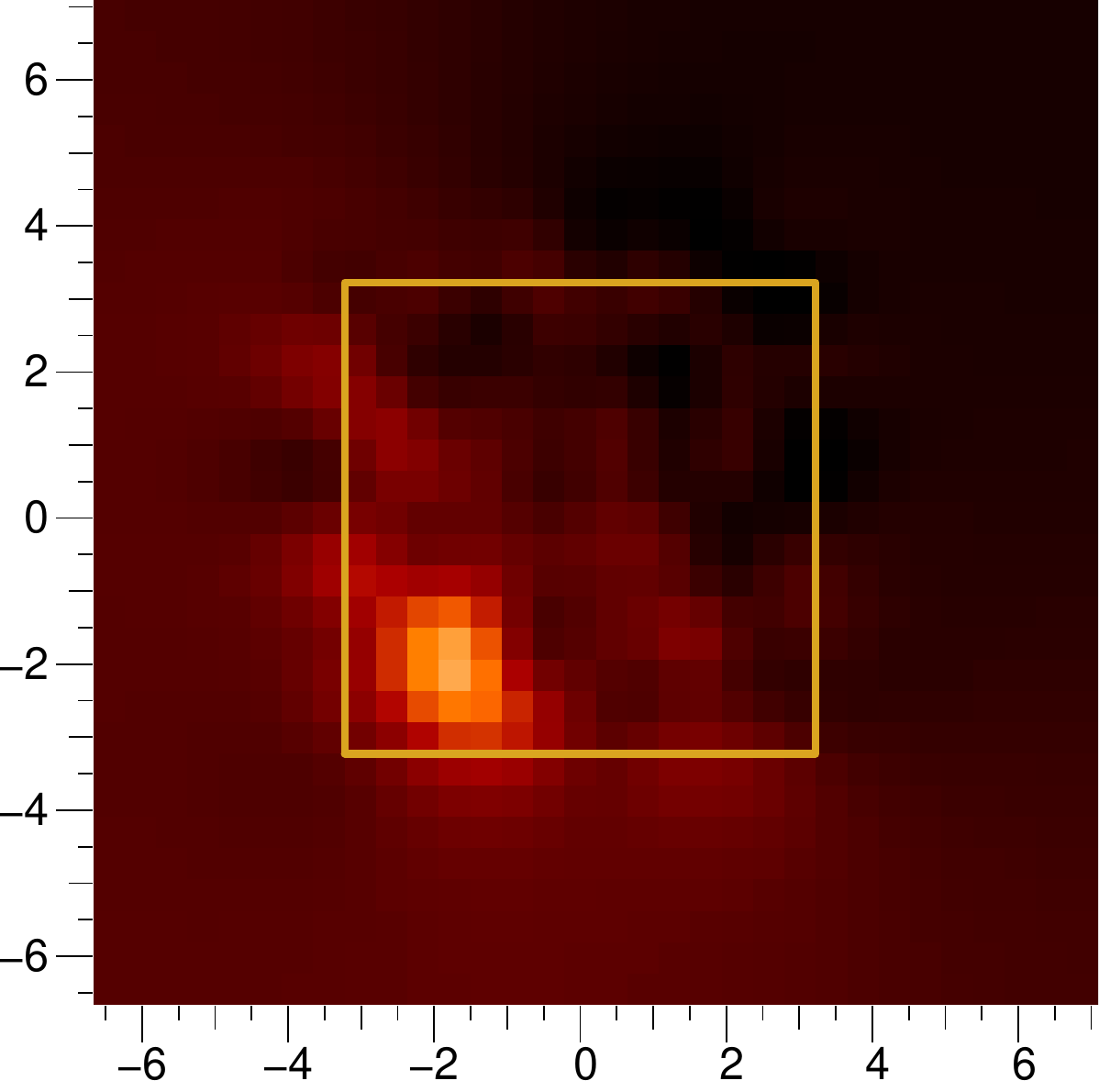} 
  \hspace{1mm}
  \includegraphics[scale=0.3]{fig_2c.pdf} 
  \caption{From left to right: reconstruction with our better regularization,
    reconstruction without any spectral regularization and reconstruction with
    spectral regularization but with constant spatial and spectral
    hyper-parameters. The three images are displayed for one single wavelength
    bin ($\lambda = 3654$\AA). The color bar is the same for all 3 figures,
    and is in relative units between the minimum and the maximum of the ground
    truth signal image displayed in figure \ref{fig:fig2}. The yellow box
    represents the data support region.}
  \label{fig:fig2bis}
\end{figure*}

Figure \ref{fig:fig2bis} illustrates the relative impact of the
regularizations. The left panel displays one wavelength slice of the
reconstruction obtained with spectral and spatial regularization including
spectral flattening, while the central panel displays the reconstruction
obtained without spectral regularization, but still with spatial
regularization including the spectral flattening. These plots show the same
wavelength slice than Figure \ref{fig:fig2}, i.e. $\lambda = 3654$\AA. The
reconstruction without spectral regularization shows clear over fitting of the
noise, apparent in the granularity of low galactic signal regions. The
$\Delta\textrm{PSNR}$ of the reconstruction with both spectral and spatial
regularization including spectral flattening is 5.2 dB, to compare to 0.1 dB
for the reconstruction without spectral regularization.

The third plot from the left of Figure \ref{fig:fig2bis} displays the
reconstruction obtained with both spatial and spectral regularizations but
without spectral flattening. It shows a slight over fitting of the low
signal regions of the galaxy compared to our best regularization, and yields a
$\Delta\textrm{PSNR}$ of 4.1 dB. The rightmost panel of Figure \ref{fig:fig2bis}
contains the true galaxy for comparison purposes.
 
\begin{figure}
  \centering

  \fbox{
  \includegraphics[width=0.4 \textwidth]{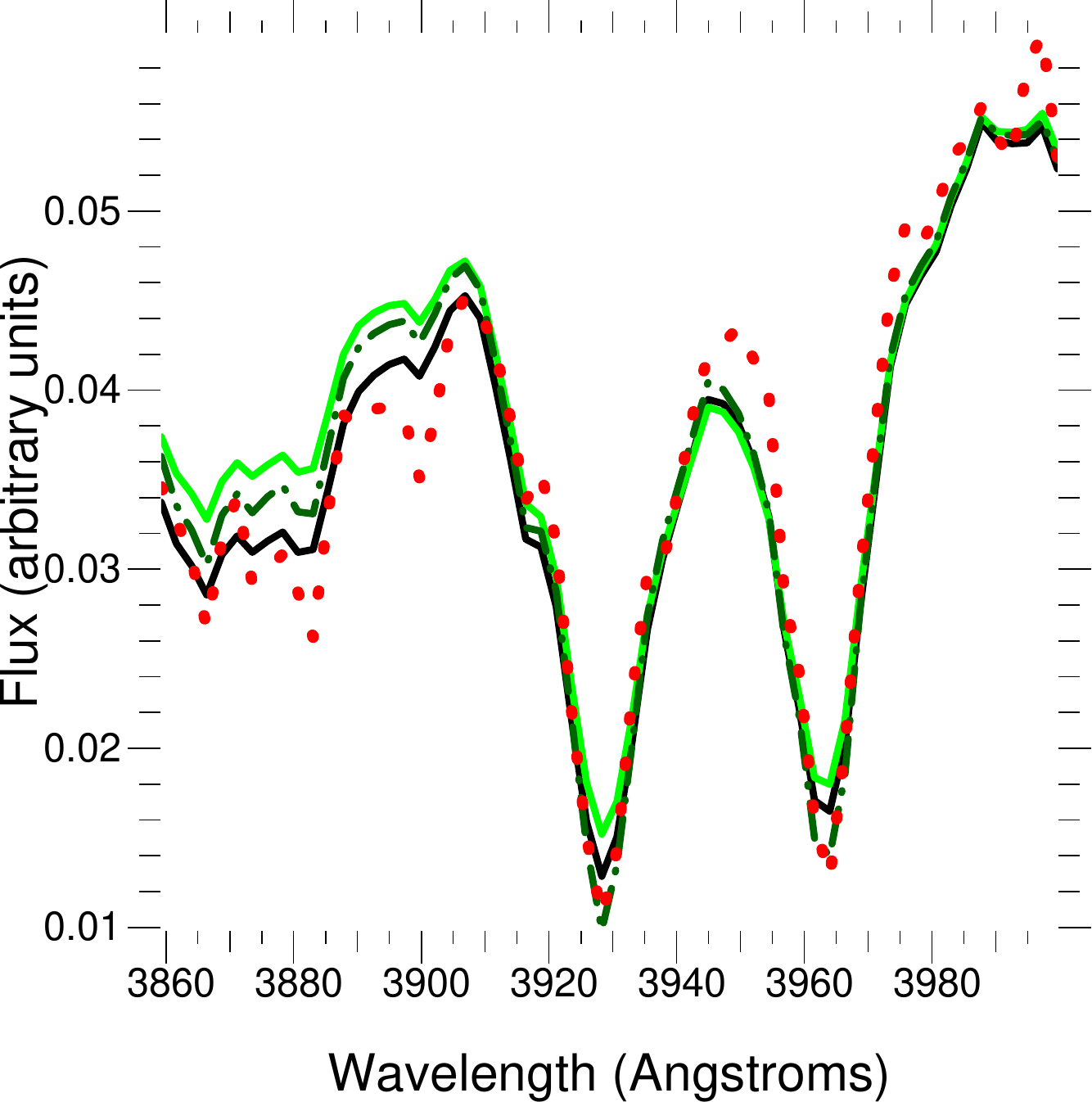}
  }
  \caption{Spectrum of the brightest spaxel of the reconstruction in black,
    superimposed over the same spaxel of the ground truth in red. We display
    in green the spectrum of the reconstruction obtained without spectral
    regularization, and in dashed dark green the spectrum of the
    reconstruction obtained with spatial and spectral regularization but
    without spectral flattening.  In order to probe the quality of the
    relative reconstruction, and in order to remove the absolute differences
    due to the difference of effective resolution between the data, the truth
    and the reconstruction, we normalize all the spectra to the same average
    value}
  \label{fig:fig3bis}
\end{figure}

\begin{figure}
  \centering
  \fbox{
  \includegraphics[width=0.4 \textwidth]{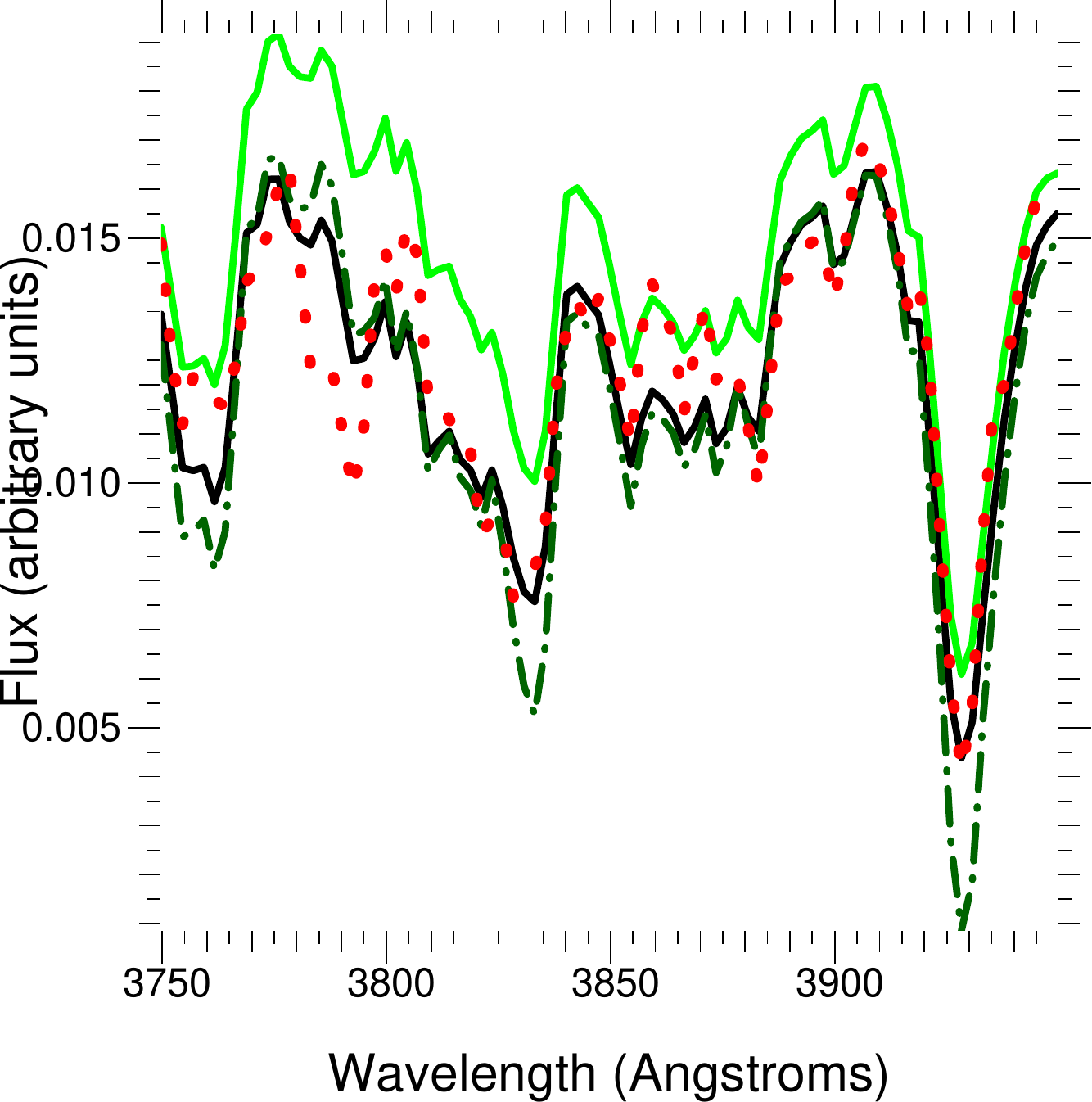}  
  }
  \caption{Spectra of the first spaxel outside of the field of view, moving
    leftwards from the brightest spaxel of the reconstruction. The
    reconstruction using the full regularization scheme is displayed in black,
    superimposed over the same spaxel of the ground truth in red. We display
    in green the spectrum of the reconstruction obtained without spectral
    regularization, and in dashed dark green the spectrum of the
    reconstruction obtained with spatial and spectral regularization but
    without spectral flattening.  In order to probe the quality of the
    relative reconstruction, and in order to remove the absolute differences
    due to the difference of effective resolution between the data, the truth
    and the reconstruction, we normalize all the spectra to the same average
    value}
  \label{fig:fig3ter}
\end{figure}

Figure \ref{fig:fig3bis} displays the spectra of the brightest spaxel of the
reconstruction for all these three reconstruction cases plus the ground truth,
zoomed on the same wavelength region than for Figure \ref{fig:fig3} bottom
plot. The black curve, corresponding to our best regularization scheme is the
closest to the ground truth, displayed in dotted red. The plain green curve,
corresponding to the reconstruction obtained without spectral regularization,
shows a consistent offset to the true spectrum: because of the lack of
spectral regularization, the fit converges toward a solution that minimizes
the spatial differences between the data and the model, to the detriment of
the spectral reconstruction accuracy. The dashed dark green curve in turn
corresponds to the reconstruction obtained without spectral flattening of the
regularizations. Even though the regularization including spectral flattening
does yield a more accurate reconstruction below 3910 \AA , the gain is maybe
less striking than in the spatial reconstruction. Figure \ref{fig:fig3ter} on
the other hand illustrates better the gain of the procedure we propose, by
showing the reconstruction accuracy of a spaxel directly outside of the field
of view. The same colors are used for the different reconstructions, and the
spaxel considered is the first spaxel outside of the field to the left (i.e
along the $Ox$ axis) of the brightest galaxy spaxel in the reconstruction. In
this case, the gain of the spectral flattening becomes more obvious: its
ability to regularize more strongly the regions with low signal allows for a
more accurate reconstruction of the spectral troughs.

\section{Results on real data}
\label{sec:data-result}

After checking the accuracy of the method on simulated data we examine the
quality of the reconstruction on real data obtained with SNIFS. To that end we
consider the host galaxy of the supernova SN2004gc discovered by The Lick
Observatory Supernova Search and followed by the supernova factory. The
exposure time of the final reference considered is 30 minutes. In order to
probe the efficiency of the reconstruction of emission lines, we concentrate on
the red channel, containing $\sim 1300$ wavelength bins and including
H$\alpha$. The time needed for the reconstruction is $\sim 10$ minutes on a
2.4 GHz cpu.

\begin{figure*}
  \centering
  \includegraphics[scale=0.4]{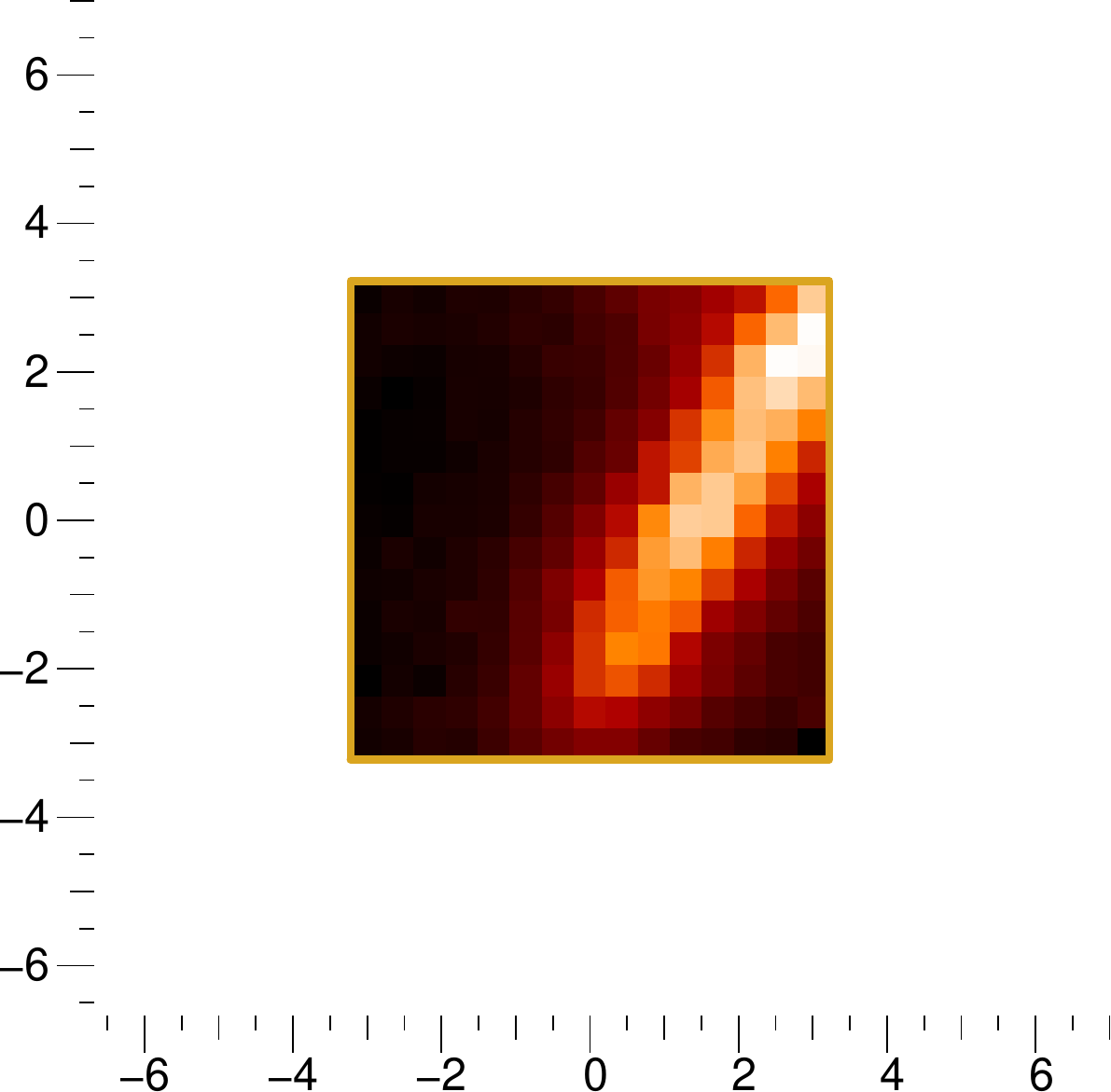}
  \hspace{1.5mm}
  \includegraphics[scale=0.4]{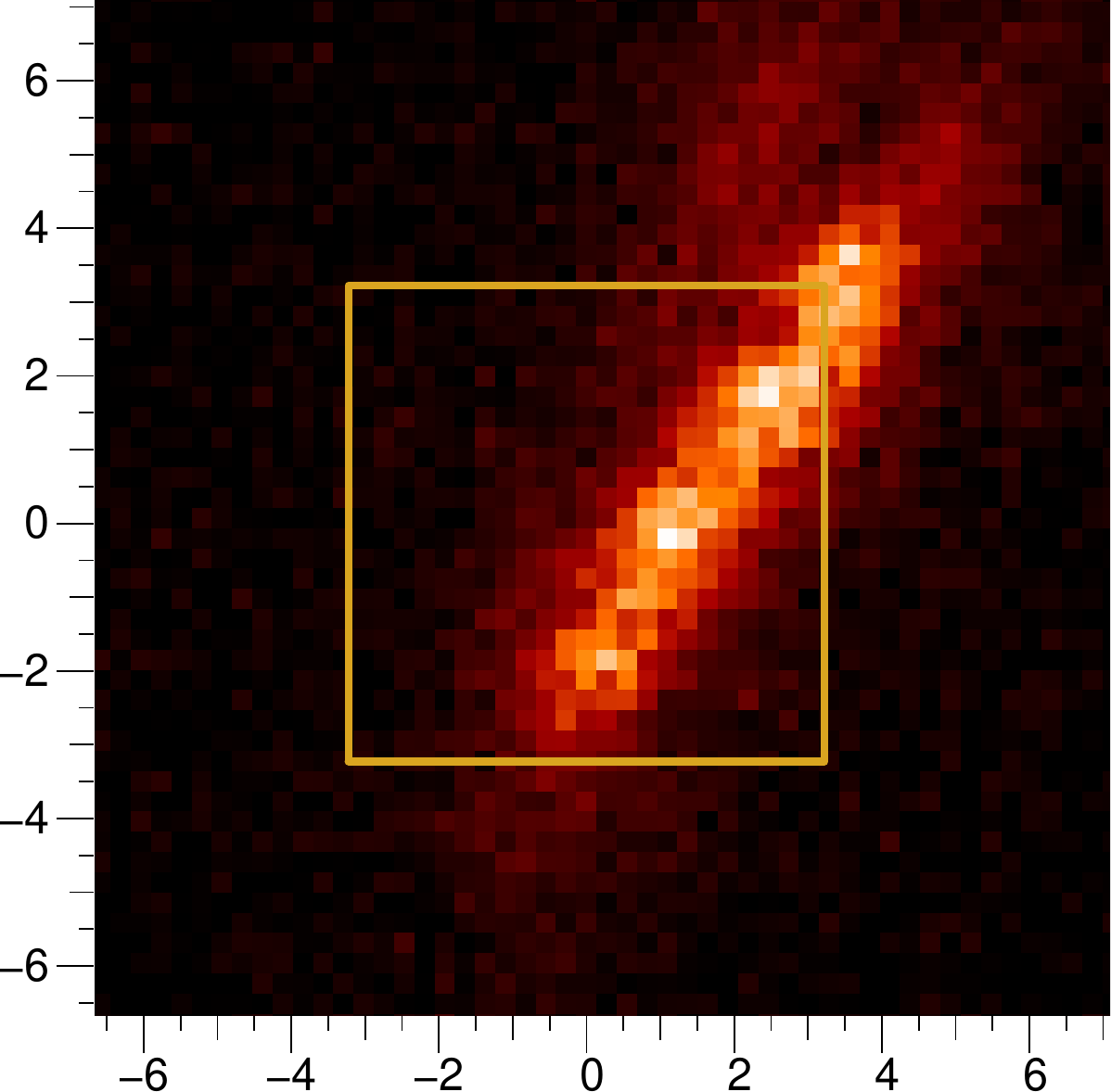}
  \hspace{1.5mm}
  \includegraphics[scale=0.4]{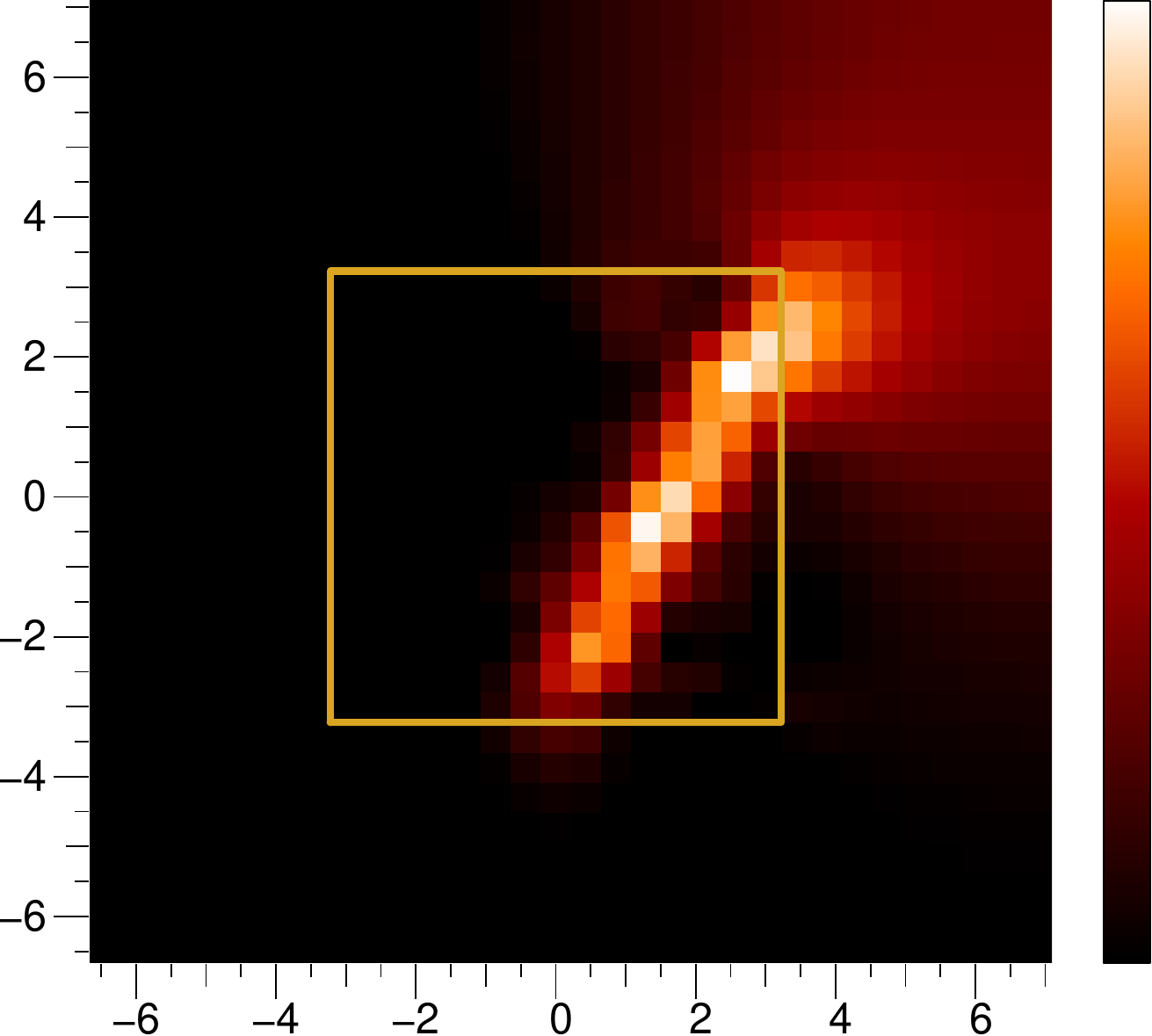}
  \caption{SN2004gc host galaxy. The left plot displays the original data, the
    central one the stack of all acquisition images obtained by SNIFS imaging
    CCD prior to each spectral exposure, and the right plot displays the
    reconstructed galaxy. Both the original data and the reconstructed galaxy
    are integrated over a top-hat synthetic V filter. Each acquisition image
    is obtained through a $V$ filter, with a 2x2 binning, yielding a pixel
    size of $\sim 0.3'' \times 0.3''$. The figure axis are labeled in
    arcseconds }
  \label{fig:fig6}
\end{figure*}

We display on the leftmost panel of Figure \ref{fig:fig6} the final reference
datacube integrated over a synthetic top-hat V filter (with a throughput of 1
between 5200 and 6289 \AA). The reconstruction is displayed on the middle
panel, with a golden box materializing the spectrograph field of view. On the
rightmost panel we display the stack of all V band acquisition images obtained by
SNIFS photometric channel prior to each spectroscopic observation. This deeper
image is on a slightly finer scale (pixel size: $0.28'' \times 0.28''$) and
observed through a V filter. In this plot also the golden box materializes SNIFS
spectroscopic field of view. Comparison of the reconstruction with the
acquisition image shows that the fine structure of the galaxy inside of the
field is well reconstructed. Moreover the field extrapolation of both the
bright core on the top right and the fainter arm on the center left of the top
end of the field of view are well extrapolated.

\begin{figure*}
  \centering
  \includegraphics[scale=0.3]{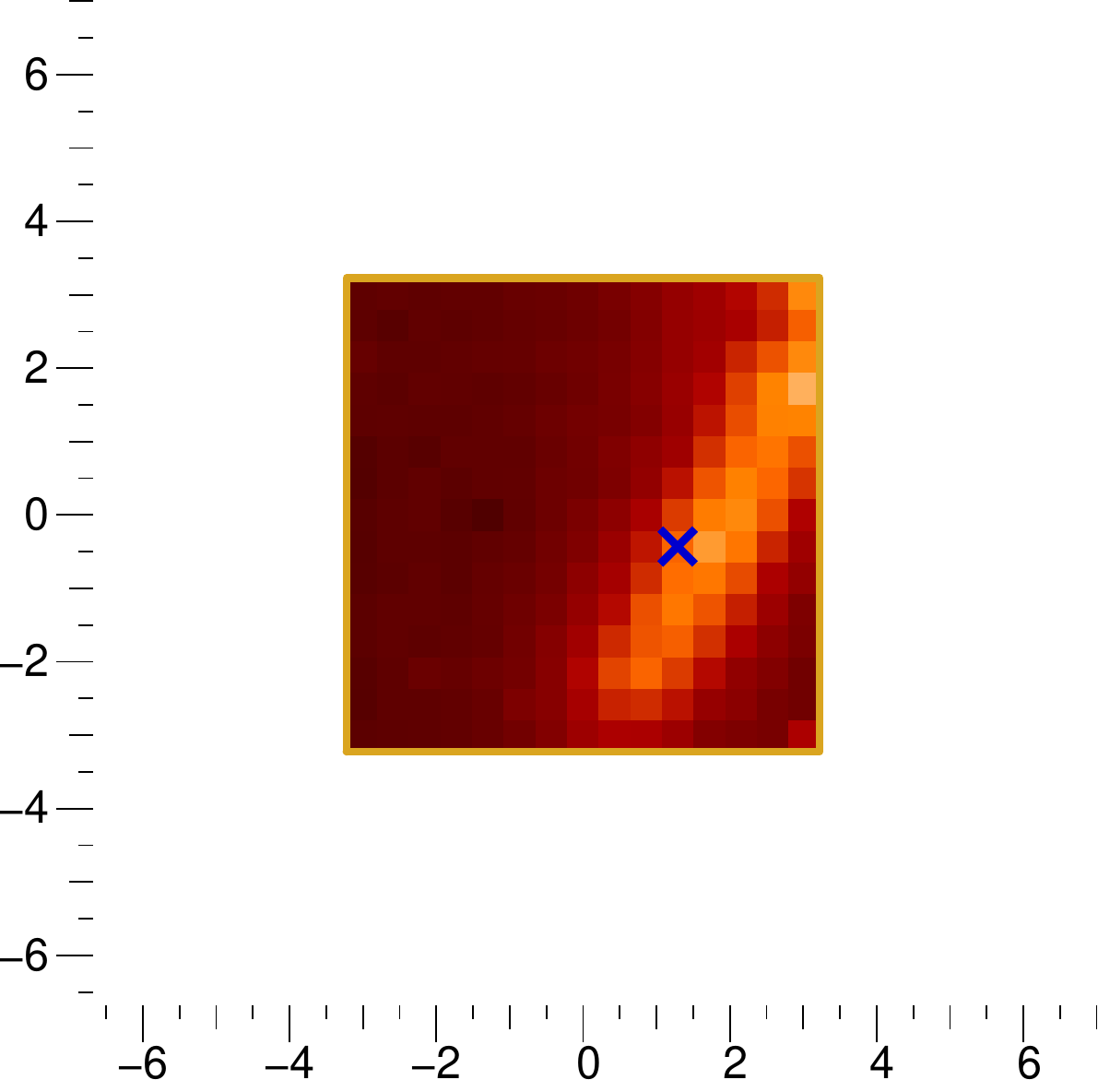}
  \hspace{1mm}
  \includegraphics[scale=0.3]{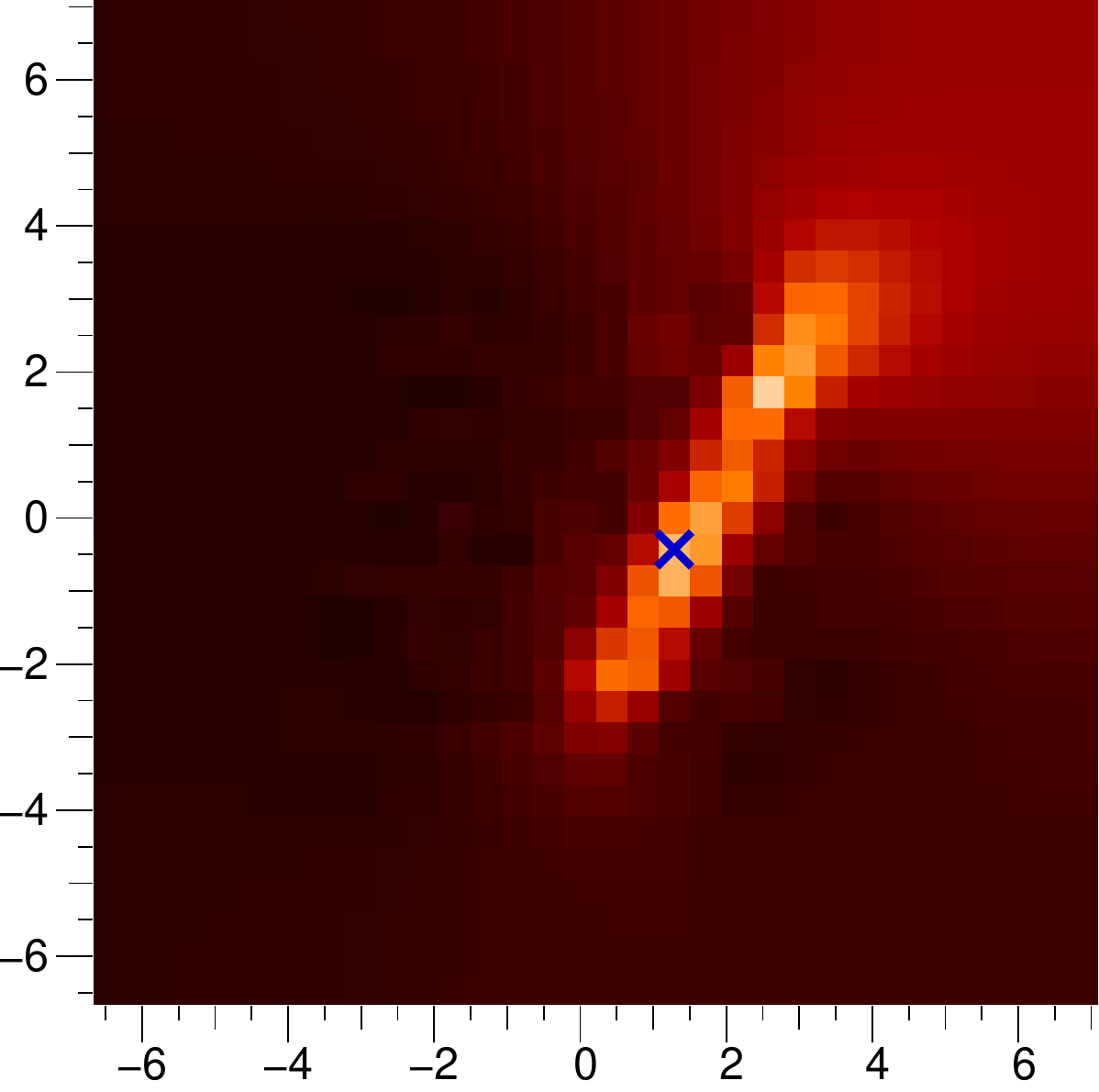}
  \hspace{1mm}
  \includegraphics[scale=0.3]{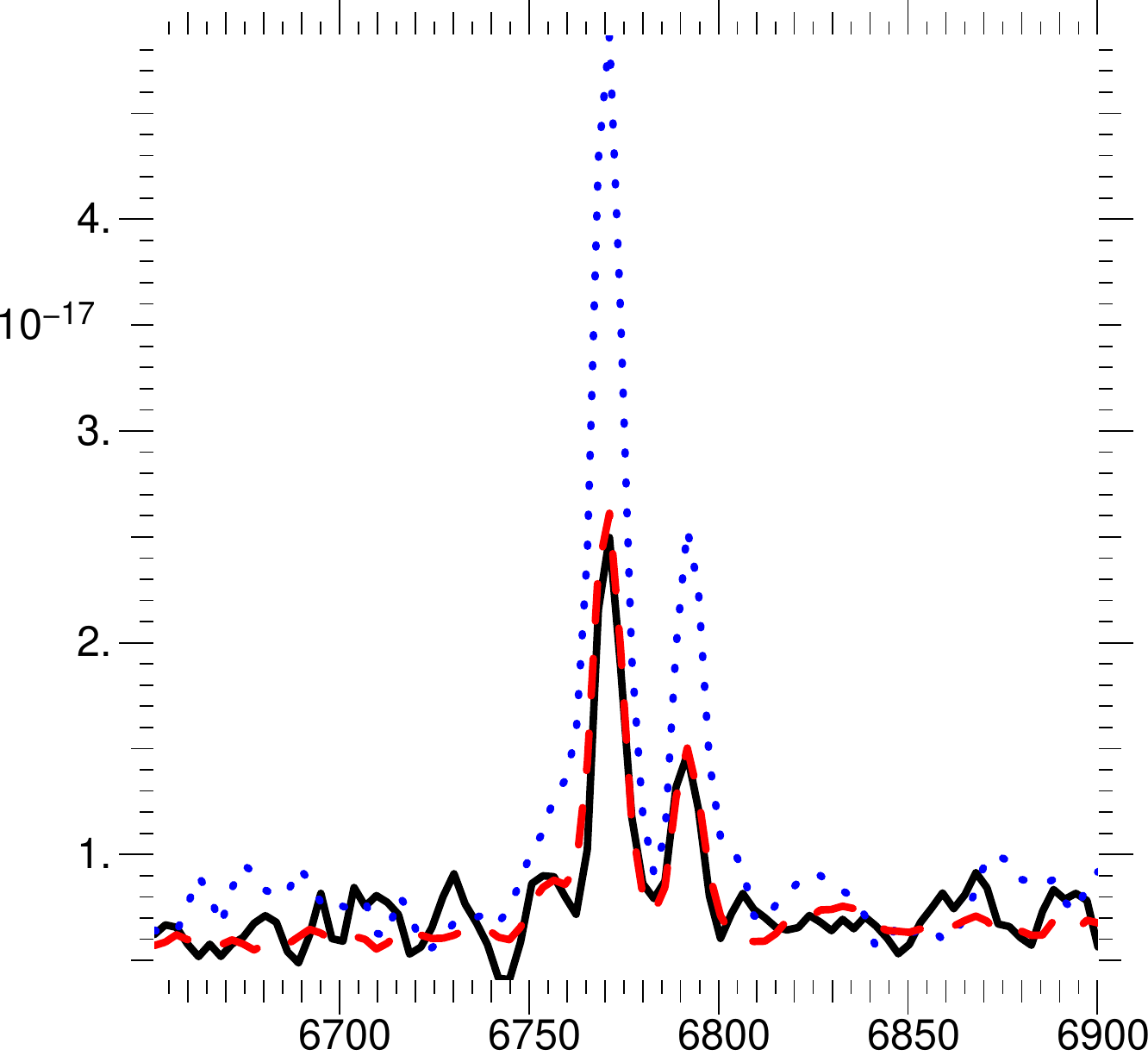}
  \hspace{1mm}
  \includegraphics[scale=0.3]{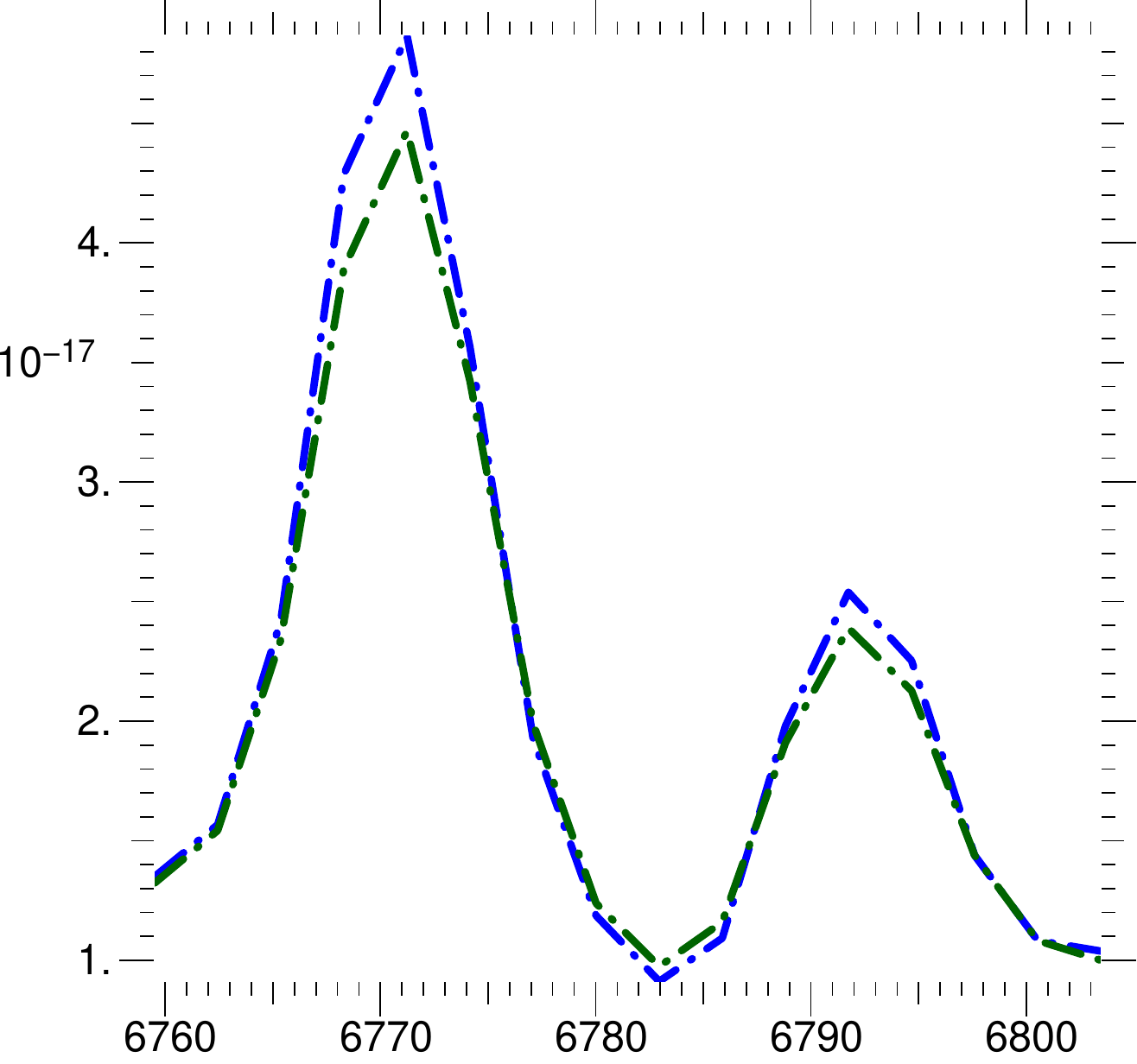}
  \caption{The two first figures from the left represent respectively the data and the
    reconstruction integrated between 6600 \AA\ and 6900 \AA. The next figure
    displays the data spectrum in black, the model in dashed red, and
    the reconstruction spectrum in dotted blue over the same wavelength range
    for the location marked by a red cross in the two first figures. Finally,
    the rightmost spectra show in dotted blue the same reconstruction than in
    the previous plot, while the dotted green spectrum corresponds to the
    reconstruction without spectral regularization and the dashed dark green
    one corresponds to the reconstruction with spatial and spectral
    regularization but without spectral flattening}
  \label{fig:fig7}
\end{figure*}

In figure \ref{fig:fig7} the same data and reconstruction are plotted on the
two leftmost panels, this time integrated over $\sim 300$\AA . The red cross
corresponds to the position of the spectrum plotted in the third panel from
the left. The spectrum of the data is displayed in black, the spectrum of the
model in dotted red and the spectrum of the reconstruction in dotted blue. The
bright region of the galaxy located under the red cross is spatially more
localized in the reconstruction, showing the effects of the spatial
deconvolution. At the same time, the flux in the data and the model at the
location of the cross are in total agreement. The strong peaked lines
displayed are enhanced by the reconstruction due to the spatial
deconvolution. The presence of larger spectral variations in the
reconstruction than in the data supports the claim that the spectral
regularization we use does not induce any pathological over smoothing in the
spectral direction. Moreover, the rightmost spectrum shows the comparative
benefits of the regularization with spectral flattening compared to that
without. The advantage of the former becomes obvious in this plot, as it
yields a better reconstruction of these strong narrow lines.

\begin{figure*}
  \centering
  \includegraphics[scale=0.3]{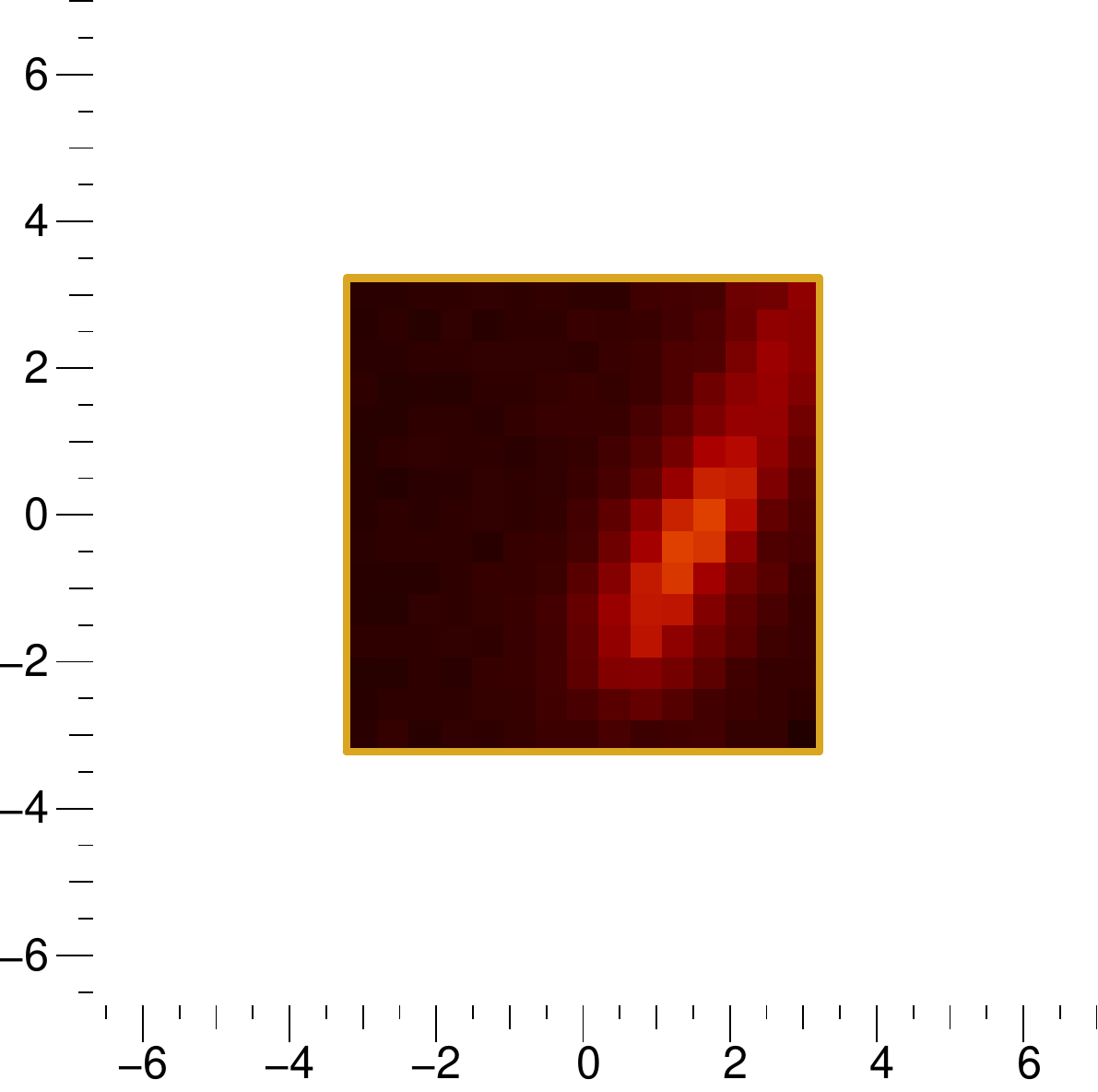}
  \hspace{1mm}
  \includegraphics[scale=0.3]{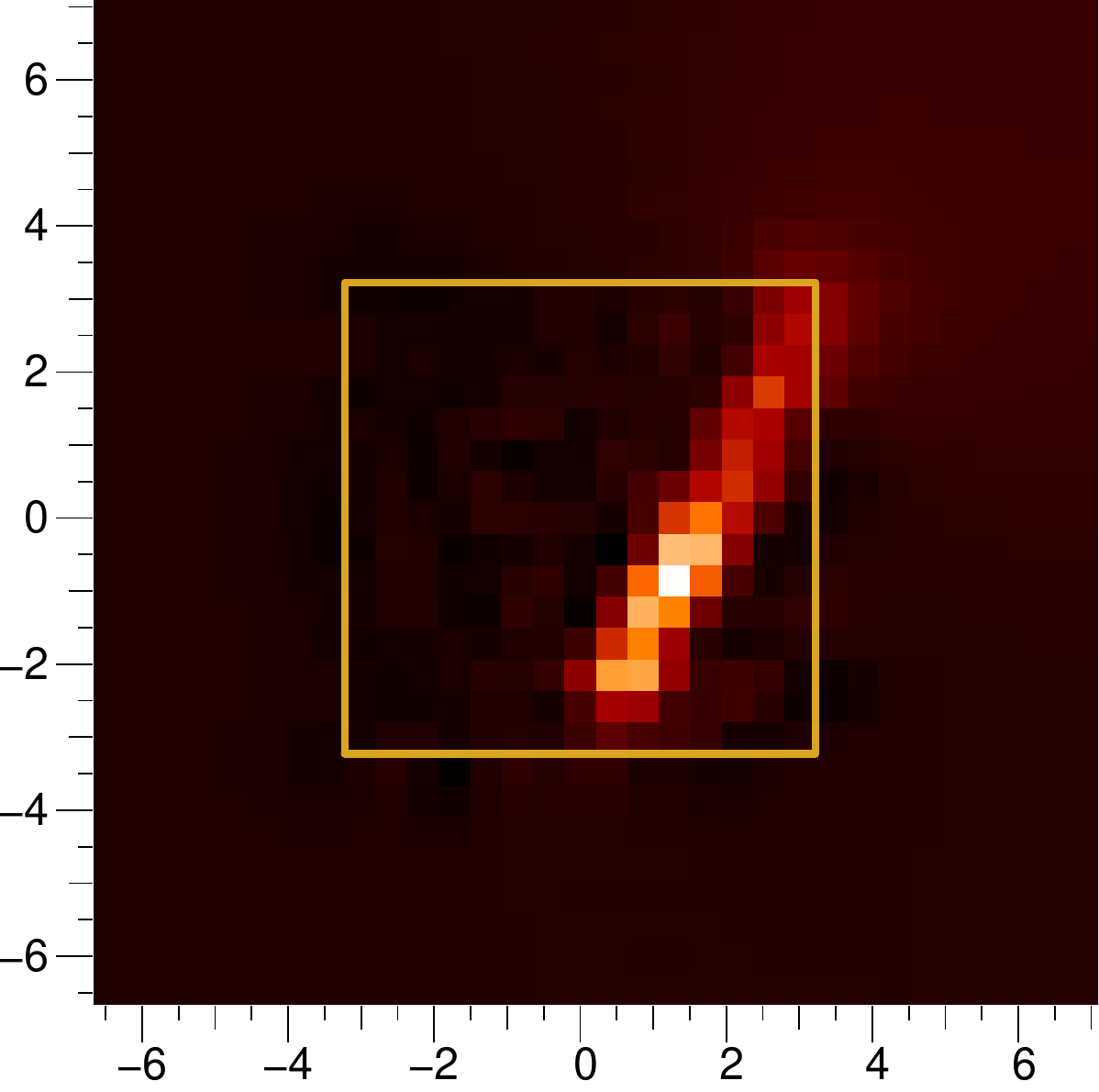}
  \hspace{1mm}
  \includegraphics[scale=0.3]{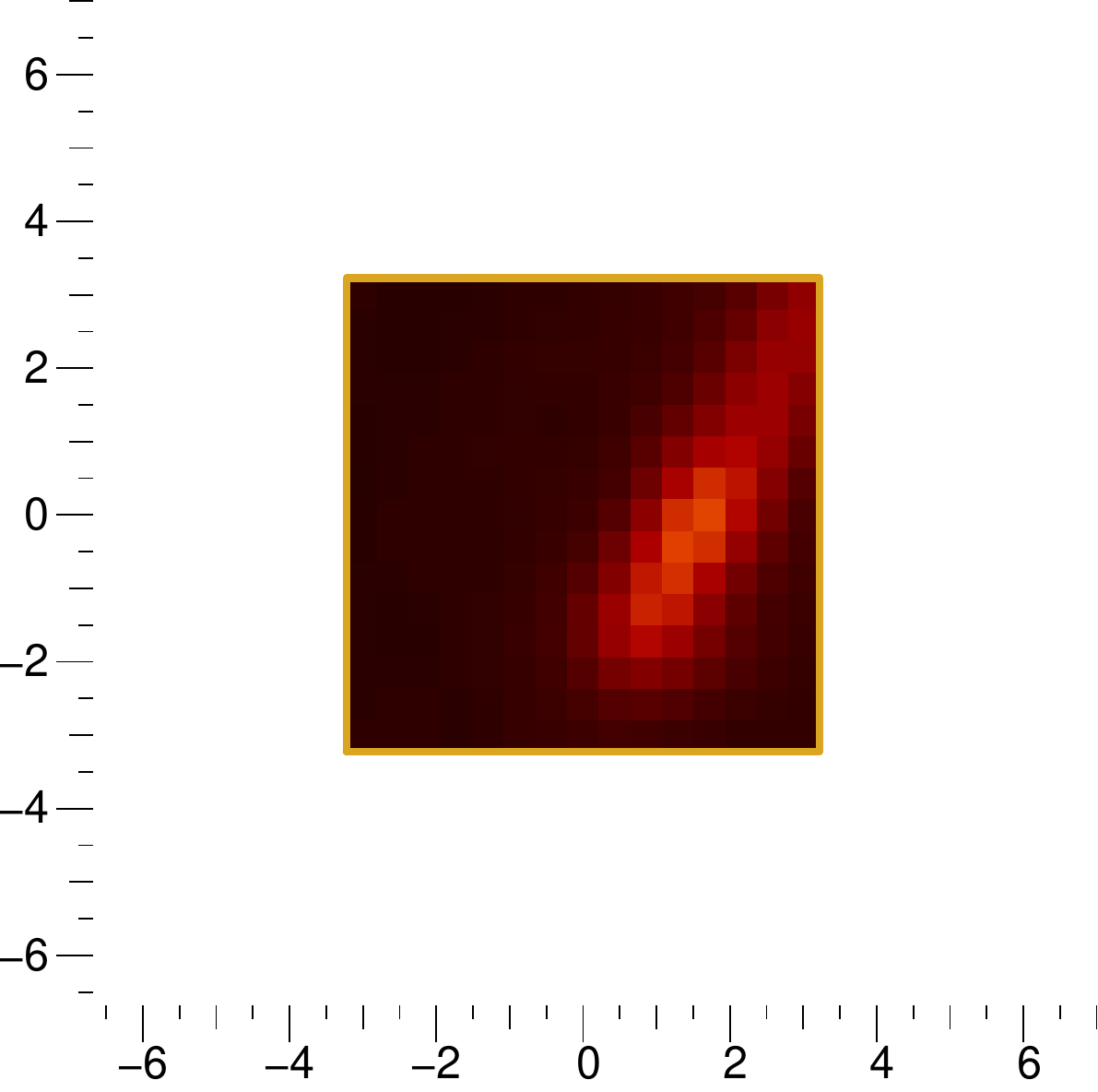}
  \hspace{1mm}
  \includegraphics[scale=0.3]{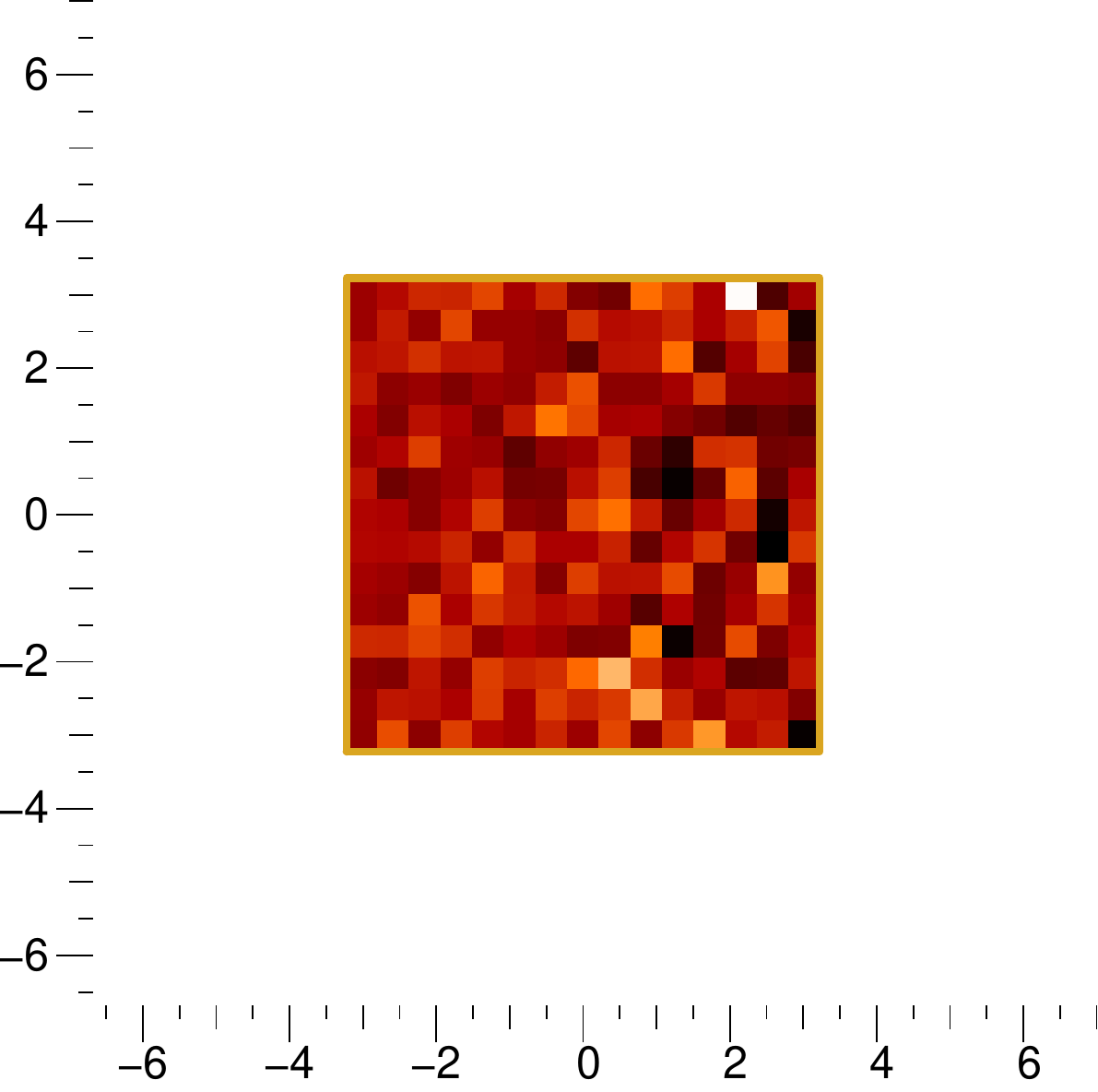}

  \caption{ The leftmost panel displays a single wavelength bin of a data cube of
    the same host galaxy obtained at a different date, and under different
    conditions than the data used for the reconstruction. The second panel
    from the left displays the reconstruction in the same wavelength bin,
    corresponding to the strong H$\alpha$ emission line shown in Figure
    \ref{fig:fig7}. Third and fourth to the right come the model of the data
    obtained with this reconstruction and the residual corresponding to the
    subtraction of this model from the data.}
  \label{fig:fig8}
\end{figure*}

The leftmost panel of Figure \ref{fig:fig8} displays only one wavelength bin
of the data, but this time corresponding to an other final reference than the
one used for the reconstruction. This second final reference has been observed
at a different epoch, with different seeing and different airmass.  The
reconstruction slice corresponding to the same wavelength is displayed on the
second panel from the left.  The next panel displays the model of the data
obtained by reconvolution of the reconstruction by the PSF corresponding to
the new exposure, and fitting of an additional flat background. The last panel
to the right  shows the residual of the subtraction of this model to the data
displayed in the first panel. The wavelength selected for this slice
corresponds to the maximum of the H$\alpha$ line displayed in the rightmost panel
of Figure \ref{fig:fig7}.

The residual shown at the right of Figure \ref{fig:fig8} is mainly
flat. Results discussed in the next section will support the claim that these
residuals are negligible with respect to the initial galactic flux. The
absence of spatially structured residual concentrated at the top right corner
of the residual image shows that the galactic light injected in the field of
view by the PSF of this observation of the host galaxy of SN2004gc is well
accounted for by the field extrapolation of our reconstruction.  Moreover, the
absence of structured residual in the rightmost plot of Figure \ref{fig:fig8}
confirms that the reconstruction bias of strong peaked features is negligible:
If that was not the case, the bright H$\alpha$ region visible at the center
right of the reconstruction slice would cause a strong localized residual for
an exposure observed under different conditions.

\section{Discussion}
\label{sec:discussion}
The method we have presented shows both good spectral and spatial accuracy in
the reconstruction of the 3-D image considered, both on simulated and real
data. Even though the quality of the reconstruction on real data can not been
checked against the ground truth, we propose here to discuss in more details
the quality of the reconstruction of SN2004gc host galaxy 3-D cube.

The reason why we selected SN2004gc as study case is because 10 different
final references have been observed over epochs spanning a large range of
different observational conditions. This allows to select one final reference
for the reconstruction and to use the 9 remaining final references to test the
quality of the model obtained after reconstruction, accounting correctly for
ADR.  Table \ref{tab:sn2004gc-apperture-photometry} shows in three synthetic
top hat filters the flux of the data and of the residual integrated spatially
over all spaxels and averaged over the 9 final references not used for the
reconstruction. For convenience, it also shows in the fifth and sixth rows the
relative value of the residual compared to the data for each of these
synthetic filters. Based on those numbers, we see that the galactic
reconstruction allows for a subtraction leaving much less than 1\% of the
initial flux in each spectral band. We checked that the error in color
was completely negligible in both V-R, R-I and V-I. Note  that the data
rms is larger than the residual average because it is dominated by the sky
variation from epoch to epoch, not by photon noise. Given that the final
references used for this comparison have been observed under a range of
observational conditions, table \ref{tab:sn2004gc-apperture-photometry} shows
that the galactic reconstruction, allows to account correctly and with
negligible bias for the galactic signal in the field of view of
SNIFS. Moreover, since the pointing variation between these exposures has an
RMS of $\sim 0.5$ spaxels, the field extrapolation can be considered to have
been tested under realistic assumptions.

\begin{table}
  \begin{center}
  \caption{  \label{tab:sn2004gc-apperture-photometry} Results of the subtraction of the reconvolution of the
    reconstructed galaxy to 9 different other final references of SN2004gc
    host obtained under a large variety of observational conditions. All three
  filters are top hat filters with bandpasses $[$5200\AA, 6289\AA$]$ for V,
  $[$6289\AA, 7607\AA$]$ for R, $[$7607\AA, 9200\AA$]$ for I.}
    \begin{tabular}{|c|c|c|c|}
      \hline
      & V filter & R filter & I filter \\
      \hline
      Data     & $9.3\,10^{{-13}}$ & $1.2\,10^{{-12}}$ & $1.9\,10^{{-12}}$ \\ 
      average  &                    & & \\
      \hline
      Data     & $3.8\,10^{{-14}}$ & $6.9\,10^{{-14}}$ & $2.2\,10^{{-13}}$ \\ 
      RMS      &                    & & \\ 
      \hline
      Residual & $2.4\,10^{{-15}}$ & $-1.9\,10^{{-15}}$ & $-4.4\,10^{{-15}}$\\ 
      average  &                    & & \\ 
      \hline
      Residual & $8.3\,10^{{-15}}$ & $2.3\,10^{{-15}}$  & $3.4\,10^{{-15}}$ \\ 
      RMS      &                    & & \\ 
      \hline
      Residual   & 0.3\%  & -0.2\% & -0.2\%\\ 
      relat. avg &  &  & \\ 
      \hline
      Residual   & 0.9\% &  0.2\% & 0.2\%\\ 
      relat. RMS &  & & \\ 
      \hline
    \end{tabular}
    \end{center}
\end{table}

\begin{figure}
  \centering
  \includegraphics[scale=0.4]{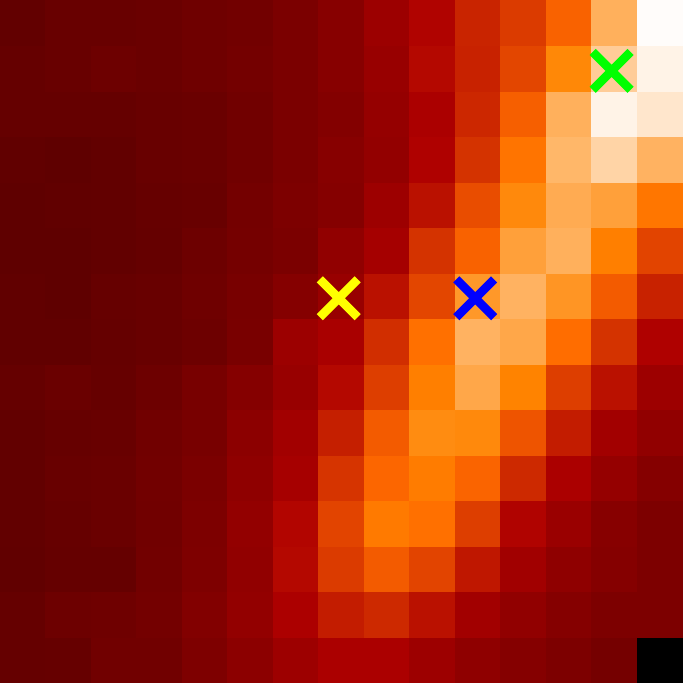}
  \caption{ The three locations used to estimate the quality of the
    subtraction as shown in tables \ref{tab:sn2004gc-psf-photometry-1} and
    \ref{tab:sn2004gc-psf-photometry-2}. The red cross corresponds to Position
  1, the blue cross to Position 2, and the green cross to Position 3.}
  \label{fig:fig9}
\end{figure}

Since for a fixed PSF the extraction of a point source flux plus a flat
background is a linear fit problem, it is easy to estimate the bias that an
additional structured background would cause. Using the 9 final references not
used for the 3-D image reconstruction allows to estimate this bias for a
set of supernova positions in the field. The three locations selected, and
shown in Figure \ref{fig:fig9}, are the true location of SN2004gc (red cross),
a bright region at the center of the host galaxy (blue cross), and a bright
region on the edge of the microlens array (green cross). The average value of
this bias as well as its rms for those three positions and two supernova
apparent V magnitudes are shown in tables \ref{tab:sn2004gc-psf-photometry-1}
and \ref{tab:sn2004gc-psf-photometry-2} under the denomination ``Before''. The
two magnitudes considered are $m_{V}=18.$ and $m_{V}=19.5$ in order to probe
the typical dynamic scale of SNfactory supernovae.

Using the final reference reconstruction, we compute a galactic
background model and subtract it to each one of the 9 additional final
references considered. Using the residuals obtained, we then estimate the
residual bias that this procedure would yield if it was to be used for host
galaxy subtraction. The average and rms values of this bias are reported in
tables \ref{tab:sn2004gc-psf-photometry-1} and
\ref{tab:sn2004gc-psf-photometry-2} under the denomination ``After''. 

For the brighter case the bias after subtraction is well below the 1\% level
for the three synthetic filters considered, and for all three of the positions
considered. The consistence of this bias value with zero for three positions
with such a different host galactic pollution support the claim that the
residual after subtraction is flat. The galactic reconstruction correctly
accounts for all the spatial structure inside of the field of view, and
extrapolates correctly the flux injected from outside of the field of view by
the PSF, for a large variety of observational conditions. This claim is
reinforced by the similar results of table
\ref{tab:sn2004gc-psf-photometry-2}, obtained for a much fainter
supernova. The bias measured is below the 1\% level on average and consistent
with zero according to the error bars.

\begin{table}
  \begin{center}
    \caption{\label{tab:sn2004gc-psf-photometry-1} Estimation of the bias on 9 different final references, before
      and after galactic subtraction in percentage of the SN flux for $m_{V,
        SN} = 18.$. This corresponds to the apparent magnitude of SN2004gc 15
      days after maximum light.}
    \begin{tabular}{|c|c|c|c|}
      \hline
      & V filter & R filter & I filter \\
      \hline
      Position 1 &  &  & \\ 
      Before & 44.6\% $\pm$ 13.0\% & 49.8\% $\pm$ 13.8\%  & 45.7\% $\pm$ 14.3\% \\ 
      After  & -0.7\% $\pm$ 0.5\% & -0.2\% $\pm$ 0.2\%  & -0.1\% $\pm$ 0.2\%\\ 
      \hline
      Position 2&  &  &  \\ 
      Before & 69.8\% $\pm$ 15.7\% & 80.8\% $\pm$ 18.0\% & 67.8\% $\pm$ 17.6\%\\ 
      After & 0.2\% $\pm$ 0.9\%  & -0.2\% $\pm$ 0.7\% & 0.3\% $\pm$ 0.3\%\\
      \hline
      Position 3 & &  &  \\ 
      Before & 61.4\% $\pm$ 11.8\% & 74.1\% $\pm$ 15.9\% & 67.7\% $\pm$ 17.2\% \\ 
      After & 0.4\% $\pm$ 0.8\%  & 0.0\% $\pm$ 0.5\% & 1.3\% $\pm$ 1.0\% \\
      \hline
    \end{tabular}
    \end{center}
\end{table}

\begin{table}
  \begin{center}
    \caption{\label{tab:sn2004gc-psf-photometry-2} Estimation of the bias on 9 different final references, before
      and after galactic subtraction in percentage of the SN flux for $m_{V,
        SN} = 19.5$. This is 0.5 magnitudes fainter than the apparent $V$
      magnitude of SN2004gc 40 days after maximum light.}
    \begin{tabular}{|c|c|c|c|}
      \hline
      & V filter & R filter & I filter \\
      \hline
      Position 1 &  &  & \\ 
      Before & 177.5\% $\pm$ 51.9\% & 198.4\% $\pm$ 55.2\%  & 181.8\% $\pm$ 56.9\% \\ 
      After  & -2.7\% $\pm$ 2.0\% & -0.0\% $\pm$ 1.0\%  & -0.3\% $\pm$ 0.8\%\\ 
      \hline
      Position 2&  &  &  \\ 
      Before & 277.7\% $\pm$ 62.4\% & 321.7\% $\pm$ 71.7\% & 270.\% $\pm$ 70.2\%\\ 
      After & 0.9\% $\pm$ 3.6\%  & -0.7\% $\pm$ 2.9\% & 1.2\% $\pm$ 1.3\%\\
      \hline
      Position 3 & &  &  \\ 
      Before & 244.5\% $\pm$ 47.0\% & 294.9\% $\pm$ 63.4\% & 269.5\% $\pm$ 68.5\% \\ 
      After & 1.8\% $\pm$ 3.4\%  & 0.2\% $\pm$ 2.0\% & 5.4\% $\pm$ 3.9\% \\
      \hline
    \end{tabular}
    \end{center}
\end{table}

It is also to be noted that the algorithm we propose here trivially allows to
calculate the galactic reconstruction using \emph{all} the final references
available simultaneously. Moreover, extending our approach to adjusting
simultaneously all supernova exposures, even those including a supernova,
opens exciting prospectives along the lines of the demixing of the supernova
signal from the other components. Even though explorations we lead in this
direction have yielded encouraging results, they are beyond the scope of the
current paper.

Applying our algorithm to real data, we notice that the two hyper-parameters
that our procedure leaves to tune are approximately the same for a range of
SuperNova factory final references, with the default values
$\mu_\Tag{spatial}=10^{-3}$ and $\mu_\Tag{spectral}=5\,10^{-3}$. This is not too
surprising since the spectral variability is accounted for in the way we deal
with the hyper-parameters. Also, in the redshift range considered, lots of the
strong galaxies we have been concentrating on have similar spatial
structures. On the other hand, more precise reconstruction can be achieved by
refining the hyper-parameter values by trial and error. Further studies are
needed to investigate the possibility of refining those values automatically:
we still have to determine whether existing hyper-parameters setting methods
(like SURE\citep{Stein1981}) could be used or whether a more specific method
has to be developed.

Another path of further studies worth mentioning concerns our use of quadratic
regularization. In this work, we choose to enforce both spatial and spectral
priors using quadratic regularization functionals in order to be able to
estimate \emph{a posteriori} covariance. However if scientific goals do not
require such posterior covariance, a possibly better reconstruction could be
achieved using non-quadratic regularization such as the edge preserving
multispectral regularization proposed by \citet{Schultz1995} or the total
variation for vector-valued images \citep{ Sapiro1996, Blomgren1998,
  Tschumperle2002}. These paths are currently under study.

\section{Conclusion}
We have seen that using the regularization scheme presented in Section
\ref{sec:regul} allows us to take advantage of the continuities present in the
data. The reconstructions obtained allow for realistic field extrapolation and
spectral reconstruction even of strong narrow lines. The field extrapolation,
as well as the quality of the reconstruction inside of the field, even in
spectral bins where the galactic signal is low, permit an accurate subtraction
of the host galaxy of supernova observed by the SuperNova Factory. For typical
SuperNova factory supernovae magnitudes, the subtraction residuals are below
the percent level for all the synthetic filters considered. Given that this
result is obtained on a set of final references intentionally obtained under
very different observational conditions, this validates the quality of the
reconstruction on real data.

Moreover, we have shown that the sharp galactic features are minimally biased
in the reconstruction process both on simulations and real data. Even though
the spatial and spectral regularization tend to bias the signal toward a
smooth reconstruction, our use of spectrally variable hyper-parameters allows
to maintain this bias below the noise level for all locations and wavelengths.

Our algorithm and the VMLM-B minimizer that we use can deal with
the large number of parameters and data points needed to simultaneously adjust
all final references obtained at once. Further work will expand in the
direction of simultaneous demixing of the supernova signal: Encouraging
results have been obtained by fitting simultaneously all exposures, including
those containing the supernova.  Another path worth exploring is that of the
blind deconvolution. Given the strong smoothness of the wavelength dependence
of the PSF, each exposure containing a supernova will yield strong constraints
on the PSF shape, and could allow for a simultaneous adjustment of the PSF,
demixing of the supernova signal from the others, and reconstruction of the
host galaxy.

Even though changes in the data format and presentation could need non trivial
implementation adjustments, the algorithm and method presented here translate
directly to any data that has multiple instances of the same object in
different spectral bands. As a consequence, this technique would be suitable
for other projects that aim at obtaining hyper-spectral data in the future.
MUSE for example \citep{Henault2003} will obtain images of $1' \times 1' $
regions of the sky from 480nm to 930nm with a spectral resolution of
$3000$. This is typically a situation in which the techniques developed here
could be applied. For example, even though the final goal of the technique we
present is unbiased host subtraction, the image restoration method could also
yield an increase in spectral and spatial resolution of the reconstructed
image. This ability would be of special interests for weak lensing surveys,
for which the measurement of the shape of the galaxy uncontaminated by the
observational blur is of critical importance.

\section*{Acknowledgments}
The authors wish to acknowledge that this work was undertaken as part of the
SuperNova Factory project, and wish to thank the entire collaboration for
fruitful discussions. The algorithm development profited strongly from its
implementation in production and its intensive application to a large quantity
of real data. We also thank the Nearby Supernova Factory collaboration for the
acquisition of multiple instances of the same host galaxy used for this study.

We are very grateful to our referee Laurent Mugnier for his thorough review and
profitable as well as constructive comments.

During a part of this work, Ferréol Soulez has been funded by the French
Agence Nationale de la Recherche on the MiTiV project (ANR-09-EMER-008-01).

Sébastien Bongard acknowledges support in part from the US Department of Energy Scientific
Discovery through Advanced Computing program under contract DE-FG02-06ER06-04,
from Director, Office of Science, Office of High Energy Physics, of the
U.S. Department of Energy under Contract No. DE-AC02-05CH11231; by a grant
from the Gordon \& Betty Moore Foundation; and in France by support from
CNRS/IN2P3. 

Our algorithms have been implemented in \textsc{Yorick}, a fast interpreted
data processing language developed by D.\ Munro and freely available at
\href{http://yorick.sourceforge.net/}{http://yorick.sourceforge.net/}.

%\appendix

\bibliographystyle{mn2e}
\bibliography{./IEEEabrv,./ddt-biblio}

\label{lastpage}

\end{document}

%% file: ddt-macros.tex
\newcommand{\OP}[1]{{\operatorname{#1}}} % upright (*) for names
\newcommand{\Tag}[1]{\mathsf{#1}}        % math font for labels and tags
\newcommand{\V}[1]{\boldsymbol{#1}}      % vector
\newcommand{\M}[1]{\mathbf{#1}}          % matrix
\newcommand{\T}{^\mathrm{T}}             % matrix/vector transpose
\newcommand{\D}[1]{{\mathrm{d}#1}}       % integrand/differential
\newcommand{\Var}{\operatorname{Var}}    % variance
\newcommand{\avg}[1]{\langle #1\rangle}

\newcommand{\expectation}[1]{\OP{E}\!\left\{#1\right\}}

\newcommand{\argmin}{\mathop{\mathrm{arg\,min}}\limits}

\newcommand{\bydef}{\stackrel{\mathrm{def}}{=}}

\newcommand{\cf}{\emph{cf.}\xspace}
\newcommand{\eg}{\emph{e.g.}\xspace}
\newcommand{\ie}{\emph{i.e.}\xspace}

\newcommand{\Eq}[1]{Eq.~(\ref{#1})}

\makeatletter

\newcommand{\@SymbolBuilderWithLabel}[4]{%
  \begingroup \escapechar\m@ne\xdef\my@tempa{\string#1}\endgroup
  \expandafter\@ifundefined{\my@tempa}{%
    \def\my@tempb{[} % to avoid mismatching below
    \expandafter\edef\csname\my@tempa\endcsname{%
      \noexpand\@ifnextchar\my@tempb%
          {\csname\my@tempa @b\endcsname}%
          {\csname\my@tempa @a\endcsname}%
    }
    \expandafter\def\csname\my@tempa @a\endcsname{%
      #2_{#4}%
    }
    \expandafter\def\csname\my@tempa @b\endcsname[##1]{%
      {#3^{#4}_{##1}}%
    }
  }{\@latex@error{\noexpand#1is already defined}\@ehc}%
}
\newcommand{\@SymbolBuilder}[3]{% see \@SymbolBuilderWithLabel for comments
  \begingroup \escapechar\m@ne\xdef\my@tempa{\string#1}\endgroup
  \expandafter\@ifundefined{\my@tempa}{%
    \def\my@tempb{[}
    \expandafter\edef\csname\my@tempa\endcsname{%
      \noexpand\@ifnextchar\my@tempb%
          {\csname\my@tempa @b\endcsname}%
          {\csname\my@tempa @a\endcsname}%
    }
    \expandafter\def\csname\my@tempa @a\endcsname{%
      #2%
    }
    \expandafter\def\csname\my@tempa @b\endcsname[##1]{%
      {#3_{##1}}%
    }
  }{\@latex@error{\noexpand#1is already defined}\@ehc}%
}

\newcommand{\SymbolWithLabel}[3]{%
  \@SymbolBuilderWithLabel{#1}{#2}{#2}{#3}}
\newcommand{\Symbol}[2]{%
  \@SymbolBuilder{#1}{#2}{#2}}

\newcommand{\VectorSymbolWithLabel}[3]{%
  \@SymbolBuilderWithLabel{#1}{\V{#2}}{#2}{#3}}
\newcommand{\VectorSymbol}[2]{%
  \@SymbolBuilder{#1}{\V{#2}}{#2}}

\newcommand{\MatrixSymbolWithLabel}[3]{%
  \@SymbolBuilderWithLabel{#1}{\M{#2}}{#2}{#3}}
\newcommand{\MatrixSymbol}[2]{%
  \@SymbolBuilder{#1}{\M{#2}}{#2}}

\makeatother

\newcommand{\Grid}[1]{#1'}
\Symbol{\Wavelength}{\lambda}
\Symbol{\GridWavelength}{\Grid{\Wavelength}}
\newcommand{\SpectralSpline}{b_\Tag{spectral}}
\newcommand{\SpectralGrid}{\mathcal{G}_\Tag{spectral}}
\newcommand{\SpatialSpline}{b_\Tag{spatial}}
\newcommand{\SpatialGrid}{\mathcal{G}_\Tag{spatial}}

\newcommand{\DirLetter}{\theta}
\Symbol{\Dir}{\V{\DirLetter}}
\Symbol{\GridDir}{\Grid{\V{\DirLetter}}}

\newcommand{\TagModel}{\Tag{model}}
\newcommand{\TagData}{\Tag{data}}
\SymbolWithLabel{\Imodel}{I}{\TagModel}
\SymbolWithLabel{\Idata}{I}{\TagData}
\newcommand{\TagObj}{\Tag{obj}}
\SymbolWithLabel{\Iobj}{I}{\TagObj}
\VectorSymbolWithLabel{\Xobj}{\ParamLetter}{\TagObj}
\SymbolWithLabel{\Xbobj}{\V{\ParamLetter}}{\TagObj}

\newcommand{\TagDark}{\Tag{dark}}
\SymbolWithLabel{\Idark}{I}{\TagDark}

\newcommand{\TagBg}{\Tag{bg}}
\SymbolWithLabel{\Ibg}{I}{\TagBg}
\VectorSymbolWithLabel{\Xbg}{\ParamLetter}{\TagBg}
\SymbolWithLabel{\Xbbg}{\V{\ParamLetter}}{\TagBg}

\newcommand{\TagGal}{\Tag{gal}}
\SymbolWithLabel{\Igal}{I}{\TagGal}
\VectorSymbolWithLabel{\Xgal}{\ParamLetter}{\TagGal}
\SymbolWithLabel{\Xbgal}{\V{\ParamLetter}}{\TagGal}

\newcommand{\TagSky}{\Tag{sky}}
\SymbolWithLabel{\Isky}{I}{\TagSky}
\VectorSymbolWithLabel{\Xsky}{\ParamLetter}{\TagSky}
\SymbolWithLabel{\Xbsky}{\V{\ParamLetter}}{\TagSky}

\newcommand{\TagSN}{\Tag{sn}}
\SymbolWithLabel{\Fsn}{F}{\TagSN}
\SymbolWithLabel{\Isn}{I}{\TagSN}
\VectorSymbolWithLabel{\Xsn}{\ParamLetter}{\TagSN}
\SymbolWithLabel{\Xbsn}{\V{\ParamLetter}}{\TagSN}

\newcommand{\ParamLetter}{x}
\VectorSymbol{\Param}{\ParamLetter}

\Symbol{\PSF}{h}
\VectorSymbolWithLabel{\Xpsf}{x}{\Tag{psf}}
\MatrixSymbol{\LinOp}{H}

\newcommand{\DataSymbol}{y}
\VectorSymbol{\Data}{\DataSymbol}

\newcommand{\ModelSymbol}{m}
\VectorSymbol{\Model}{\ModelSymbol}

\newcommand{\ErrorSymbol}{e}
\VectorSymbol{\Error}{\ErrorSymbol}

\newcommand{\ResidualLetter}{r}
\VectorSymbol{\Residual}{\ResidualLetter}

\MatrixSymbolWithLabel{\Cerror}{C}{\Tag{err}}
\MatrixSymbolWithLabel{\Werror}{W}{\Tag{err}}
\MatrixSymbolWithLabel{\Cprior}{C}{\Tag{prior}}
\MatrixSymbolWithLabel{\Wprior}{W}{\Tag{prior}}

\newcommand{\Fcost}{f}
\newcommand{\Fdata}{\Fcost_\Tag{data}}
\newcommand{\Fprior}{\Fcost_\Tag{prior}}

\newcommand{\Nexp}{N_t}

\newcommand{\Nfov}{N_{\FOV}}
\newcommand{\Nwave}{N_{\Wavelength}}

\newcommand{\FOV}{\Omega}

\Symbol{FDif}{\ensuremath{\textrm{D}}}
\Symbol{FDifDif}{\ensuremath{\textrm{Q}}}

\Symbol{hyper}{\ensuremath{\mu}}